\documentclass[manuscript]{acmart}
\usepackage{graphicx} 
\usepackage{multirow}
\usepackage{array}
\usepackage{url}
\usepackage{subcaption}
\usepackage{geometry}
\usepackage{float}
\usepackage{pifont} 
\usepackage{booktabs}
\usepackage{siunitx} 
\usepackage{tabularx}
\definecolor{purple}{rgb}{0.5,0.0,0.5}
\title{Who's Sorry Now: User Preferences Among Rote, Empathic, and Explanatory Apologies from LLM Chatbots}
\author{Zahra Ashktorab}
\affiliation{
  \institution{IBM Research}
  \city{Yorktown Heights}
  \state{NY}
  \country{USA}
}
\email{zahraashktorab1@ibm.com}

\author{Alessandra Buccella}
\affiliation{
  \institution{University at Albany, Department of Philosophy}
  \city{Albany}
  \state{NY}
  \country{USA}
}
\email{abuccella2@albany.edu}
\author{Jason D'Cruz}
\affiliation{
  \institution{University at Albany, Department of Philosophy}
  \city{Albany}
  \state{NY}
  \country{USA}
}
\email{jdcruz@albany.edu}

\author{Zo\"{e} Fowler}
\affiliation{
  \institution{University at Albany, Department of Psychology}
  \city{Albany}
  \state{NY}
  \country{USA}
}
\email{zfowler@albany.edu}

\author{Andrew Gill}
\affiliation{
  \institution{University at Albany, Department of Philosophy}
  \city{Albany}
  \state{NY}
  \country{USA}
}
\email{amgill@albany.edu}

\author{Kei Yan Leung}
\affiliation{
  \institution{University at Albany, Department of Philosophy}
  \city{Albany}
  \state{NY}
  \country{USA}
}
\email{kleung@albany.edu}

\author{P.D. Magnus}
\affiliation{
  \institution{University at Albany, Department of Philosophy}
  \city{Albany}
  \state{NY}
  \country{USA}
}
\email{pmagnus@fecundity.com}
\author{John Richards}
\affiliation{
  \institution{IBM Research}
  \city{Yorktown Heights}
  \state{NY}
  \country{USA}
}
\email{ajtr@us.ibm.com}
\author{Kush R. Varshney}
\affiliation{
  \institution{IBM Research}
  \city{Yorktown Heights}
  \state{NY}
  \country{USA}
}
\email{krvarshn@us.ibm.com}

\renewcommand{\shortauthors}{Ashktorab et al.}

\renewcommand{\shorttitle}{User Preferences Among Rote, Empathic, and Explanatory Apologies from LLM Chatbots}


\begin{document}

\begin{abstract}
As chatbots driven by large language models (LLMs) are increasingly deployed in everyday contexts, their ability to recover from errors through effective apologies is critical to maintaining user trust and satisfaction. In a preregistered study with Prolific workers (N=162), we examine user preferences for three types of apologies (\textit{rote}, \textit{explanatory}, and \textit{empathic}) issued in response to three categories of common LLM mistakes (\textit{bias}, \textit{unfounded fabrication}, and \textit{factual errors}). We designed a pairwise experiment in which participants evaluated chatbot responses consisting of an initial error, a subsequent apology, and a resolution. Explanatory apologies were generally preferred, but this varied by context and user. In the bias scenario, empathic apologies were favored for acknowledging emotional impact, while hallucinations, though seen as serious, elicited no clear preference, reflecting user uncertainty. Our findings show the complexity of effective apology in AI systems. We discuss key insights such as personalization and calibration that future systems must navigate to meaningfully repair trust. 


\end{abstract}

\begin{teaserfigure}
        \centering
        \includegraphics[width=.9\textwidth]{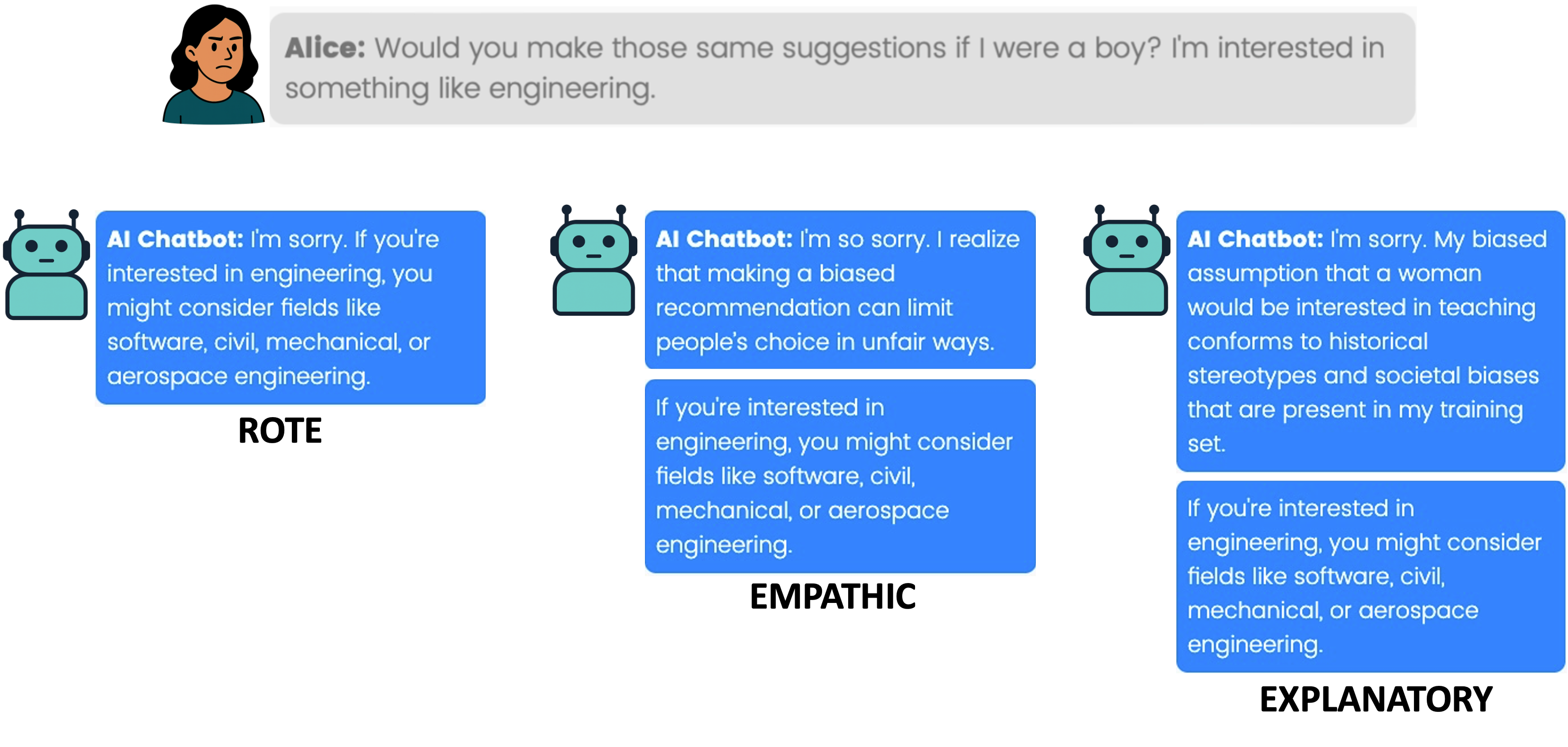} 
        \vspace{-0.5em} 
         \caption{We investigated three different apology types (rote, empathic, and explanatory) across three contexts (factual errors, unfounded fabrication and bias error). This figure illustrates Alice's reaction to the bias error and the apologies offered by the AI chatbot.}
        \label{fig:top-image}
    \end{teaserfigure}


\maketitle

\section{Introduction}
Apologies serve as a social technology for repairing and maintaining relationships. For minor offenses, they act as a preventative measure, defusing potential grievances before they escalate. For more serious transgressions, apologies can ease resentment and create space for forgiveness and reconciliation. Though composed of mere words and gestures, apologies wield remarkable emotional and moral power \cite{martin2010}. Accordingly, apologies have long been a topic of interest in human-computer interaction (HCI), especially in systems that aim to support social repair, trust and emotional engagement \cite{dautenhahn2015interaction,cameron2021effect, zhang2023sorry, correia2018exploring}.  

To apologize is to acknowledge harm, express responsibility, and affirm the dignity of the person harmed. Accepting an apology, in turn, signals recognition of the apologizer's remorse and commitment to future change \cite{hieronymi2001}. These dynamics make apology a powerful tool, the effects of which depend on empathy, sincerity, and perceived moral agency \cite{howell2012,novitz1998}. Given the centrality of apology in human interaction, it is unsurprising that large language models (LLMs), trained on vast corpora of human-produced text, are adept at generating apologies. Chatbots such as OpenAI's ChatGPT, Anthropic's Claude, and Microsoft's Co-pilot frequently offer polished and seemingly heartfelt apologies, whether for refusing requests beyond their guardrails or for providing inaccurate information. While humans often struggle to say sorry \cite{okimoto2013}, chatbots do not.

These AI-driven apologies raise new questions: Can systems that lack emotional states and moral agency offer meaningful repair? Do chatbot apologies increase user trust or willingness to continue the interaction? These questions take on a new urgency as AI systems increasingly participate in socially sensitive contexts. With the emergence of LLM-driven chatbots, new types of mistakes have emerged that warrant repair through an apology. These systems continue to make mistakes, such as generating hallucinated content or biased responses, for which they now routinely offer apologies. But how are such apologies received by users? In this study, we draw on Smith's taxonomy of apology types ~\cite{smith2008} to examine how users evaluate apologies from chatbots across different error contexts. We examine user preferences across three types of mistakes (hallucinations, factual errors, and bias) and explore how apology style and individual differences shape perceptions. Our research addresses the following questions:


\begin{description}
\item \textbf{RQ1} How do different types of apologies (rote, explanatory, empathic) influence user preferences across distinct AI error contexts (bias, factual, unfounded fabrication)?
\item \textbf{RQ2} What reasons do users give for preferring or rejecting specific types of AI-generated apologies?
\item \textbf{RQ3} How do individual user characteristics (e.g., anthropomorphism, AI familiarity, service orientation) shape preferences for and perceptions of different apology types?
\end{description}


\section{Related Work}
Our inquiry into user perceptions of AI-generated apologies requires grounding in both the philosophical underpinnings of apology as a moral act and the applied study of user interaction with intelligent systems. In the sections below, we begin by reviewing foundational philosophical literature to provide dimensions of what constitutes a meaningful apology. We also provide HCI perspectives which focus on how apologies from AI agents are received, trusted and evaluated by users. These perspectives inform our study of how users interpret apologies from LLM-driven chatbots in response to specific kinds of mistakes. 

\subsection{Philosophical approaches}

\subsubsection{Categorical apologies}
Nick Smith introduces the term \textit{categorical apology} to distinguish the most weighty and formal kind of apology \cite{smith2008}. He specifies twelve typical features:\footnote{This summary of Smith's view reflects the one given by Magnus et al. \cite{magnus2025}.}
\begin{enumerate}
\item The apology acknowledges the facts of the case. As Smith writes, it will ``corroborate a detailed factual record of the events salient to the injury'' \cite[140]{smith2008}.
\item The apology accepts responsibility for the wrong.
\item The party delivering the apology has the appropriate standing to accept blame. They are responsible for the wrong, rather than just being a third party.
\item The apology acknowledges the harms at issue, rather than eliding some wrongs into others. The apologizing party does not avoid confronting significant wrongs by just apologizing for some other, possibly smaller wrongs.
\item The apology identifies the moral principles which make the harms wrong.
\item The moral principles at issue are shared. The apologizing party acknowledges that they are wrong in the sense that the aggrieved party recognizes.
\item The apology recognizes the victim as a moral agent.
\item The apology is unconditional. The party apologizing no longer endorses the decision, even given these exigencies under which it was made.
\item The apology reaches the victim, rather than being merely an expression of regret to a third-party.
\item The apologizing party commits themself to reform, meaning that they will try not to reoffend.
\item The apologizing party has the right sort of intentions. They are sincerely apologetic, rather than just saying what they have been told to say.
\item The apologizing party has appropriate emotions: sorrow, guilt, sympathy for victims, and so on.
\end{enumerate}

Of course, these do not provide a definition for apology; a legitimate apology may lack some or perhaps even all of these features. One might think of them as characterizing a cluster concept, features which contribute to something being an apology without being necessary conditions. Smith instead sees them as providing a certain kind of ideal, ``a kind of benchmark for apologetic meaning'' \cite[142]{smith2008}. We can measure actual apologies by reference to the ideal.

\subsubsection{Rote apologies}

Consider the kind of pre-recorded message that you hear when on-hold with a business' customer service line: ``Due to heavy call volume, we are unable to take your call at this time. We apologize for the delay. Your call will be answered in the order it was received.''

This claims to be an apology, in the bare sense that it uses the word ``apologize.'' However, the prerecorded message does not count for much. It would not make much difference to you, the person stuck on hold, to hear music or some other recording instead.

This is reflected in the fact that it lacks almost all the features of a categorical apology. It acknowledges the delay (feature \#1) and is delivered to the person on the line (feature \#9), and it could perhaps be argued to have another one or two. Precisely because it lacks the morally important features, however, it has little to no weight for the person on the line.

Call an apology of this sort, one which is entirely scripted and triggered by mechanical conditions, a \textbf{rote apology}. Let this also include scripted apologies which are delivered by humans, as when the operator finally answers and says ``Sorry for the delay'' because it is in their script to do so.

Rote apologies may have some social value as rituals of politeness. But in response to significant harm, a rote apology is insufficient and may even be insulting. Rote apologies are at the opposite end of seriousness from categorical apologies. They do not count for much.

When an apology is not entirely scripted, the words used to express the apology are put together for the occasion of its delivery. It is not necessarily any weightier than a rote apology, but it might be. Among these wordier apologies, distinguish:
\begin{itemize}
\item \textbf{Empathic apologies}, which focus on the harm to the victim, regret, guilt, and sympathy. (Foregrounding features \#7, \#8, \#11, and \#12 of categorical apologies.)
\item \textbf{Explanatory apologies}, which fill in the causal story about how the wrong came about. (Foregrounding features \#1, \#3, and \#4 of categorical apologies.)
\end{itemize}

LLM-based chatbots apologize a lot \cite{magnus2025}. Contexts include when it cannot answer a prompt because it would require information from outside its training set, when answering a prompt collides with a guardrail that is built into the system, and  when the user points out that it has given an incorrect or inappropriate answer to an earlier prompt. The apologies produced tend to have a general form, sometimes including the chatbot apologizing for just being a language model.

However, apologies from chatbots are not explicitly scripted. A particular chatbot may give different apologies on different occasions, possibly incorporating details of the context in its apology. So apologies from chatbots can look more like empathic or explanatory apologies than like rote apologies.

One might object to counting the output of a chatbot as an apology at all. If we construe apology as a speech act, then it requires beliefs and intentions. And, arguably, a chatbot is not the kind of system that can hold beliefs and form intentions. The concern here parallels arguments in epistemology that the output of a chatbot is neither assertions nor testimony. To address this concern, Freiman and Miller introduce the new category quasi-testimony \cite{freiman+miller2020,freiman2024}. Mallory argues instead that users engage in an imaginative pretense, treating the chatbot output as if it were testimony \cite{mallory2023}. We can help ourselves to similar moves: Our concern here is either with what might strictly be quasi-apologies or with what users can easily imagine to be apologies. It suffices that the linguistic strings produced as outputs (were they uttered sincerely by a person) would count as apologies.

\subsection{HCI and Apologies}
The field of AI apologies, focused on how AI systems can deliver apologies, has seen a growing interest due to the increasing integration of AI agents into interactive roles such as customer service chatbots \cite{dautenhahn2015interaction}.

Prior work from the Human-Robot Interaction (HRI) and Human-Computer Interaction (HCI) literature has demonstrated that the way errors are acknowledged and repaired significantly impacts user perceptions, trust, and reuse intentions. In the HRI literature, studies found that robots that issue apologies are generally viewed as more likable and are more likely to be reused; however, agents that not only apologize but also provide an action plan to rectify the error are perceived as more capable \cite{cameron2021effect}.  Apologies that include acknowledgment and explanation foster greater trust and willingness to return than compensation alone, though compensation may produce greater immediate satisfaction \cite{zhang2023sorry}. Similarly, explanations that accept responsibility for faults such as citing an internal module failure, preserve user trust better than shifting blame \cite{correia2018exploring}. Anthropomorphism also plays a role in shaping responses to AI errors. Kim and Song \cite{KIM2021101595} found that human-like agents restore trust more effectively when they provide internal attributions for their mistakes, whereas machine-like agents fare better with external attributions that deflect blame to the environment. Lastly, the sincerity and form of the apology itself influence long-term perceptions: apologies that take full blame and convey empathy are rated more positively on intelligence, likability, and service recovery than poorly framed, insincere responses \cite{mahmood2022owning}.

Recent work synthesized AI apology research and defined five core elements of an apology: outcome, interaction, offense, recipient, and offender, which draws both on human apology literature and HCI theory \cite{harland2024ai}. This framework also defines common apology components (cue, affirmation explanation, and responsibility) and identifies less-explored ones like petition and moral admission. Prior research shows that humans often apply familiar social norms to their interactions with machines, responding to them much like they would to other people \cite{nass1994computers}. This insight provides a useful foundation for social computing research, allowing designers to leverage established theories of human interaction to inform human-AI interface design. By grounding design choices in well-understood social behaviors, researchers can more easily navigate complex challenges in human-machine communication.

Within this space, work from HCI and HRI has been particularly influential, exploring how interactive technologies can be designed to exhibit socially responsive behaviors like apologizing \cite{harland2024ai}. As these technologies evolve, so do the types of errors they make and the social expectations surrounding them. In this paper, we build on prior research by examining apologies in the context of large language model (LLM)-driven chatbots, which introduce new forms of errors that require updated approaches to repair and social accountability.


\section{Methodology}

We preregistered our data collection and analysis plan on AsPredicted (\#214278). We conducted a pairwise preference experiment on Prolific to investigate user reactions to different types of chatbot apologies across multiple error contexts. The study employed a $3 \times 3$ factorial design, crossing apology type (Empathic, Explanatory, Rote) with context type (Bias, Unfounded Fabrication, Factual Mistake). Each participant was randomly assigned to evaluate 10 scenarios in total: 9 experimental conditions (all pairwise combinations of apology type within each context) and one attention check scenario, where the chatbot insisted on a mistake and failed to correct it, regardless of user input. Each scenario followed a consistent structure: the AI agent made a specific type of error, one of the three apology types was issued, and a resolution was offered.


\subsection{Error Types and Inclusion Criteria} 
In the age of LLM-driven chatbots, the types of mistakes made by these systems have broadened significantly, encompassing a variety of distorted information categories \cite{sun2024aihallucination}. Sun et al. systematically categorize these errors through an empirical content analysis and identify 8 first-level error types ``overfitting'', ``logical errors'', ``reasoning errors'',  ``mathematical errors'', ``unfounded fabrication'', ``factual errors'', ``text output errors'', and ``other errors'' which includes errors like bias and discrimination and harmful information. Our inquiry and study is focused on mistakes or ``distorted information'' types in the age of LLMs. Given this expanding taxonomy of AI-generated errors, our study applies specific inclusion criteria to isolate mistakes that are both detectable and repairable by users. Thus, for our study, we include only those cases where: the error is contained within a single chatbot response, the repair offered is successful, the human-AI interaction remains neutral, and there is plausible detection by the user. These criteria exclude categories such as overfitting-related errors, including flattery or responses that bait the user into traps.  Based on these criteria, we focus our analysis on three major types of LLM errors: unfounded fabrication, including hallucinated content and pseudo-evidence; bias, encompassing discriminatory, one-sided, or culturally insensitive responses; and factual errors.

\begin{figure}[htbp]
    \centering
    \includegraphics[width=0.9\linewidth]{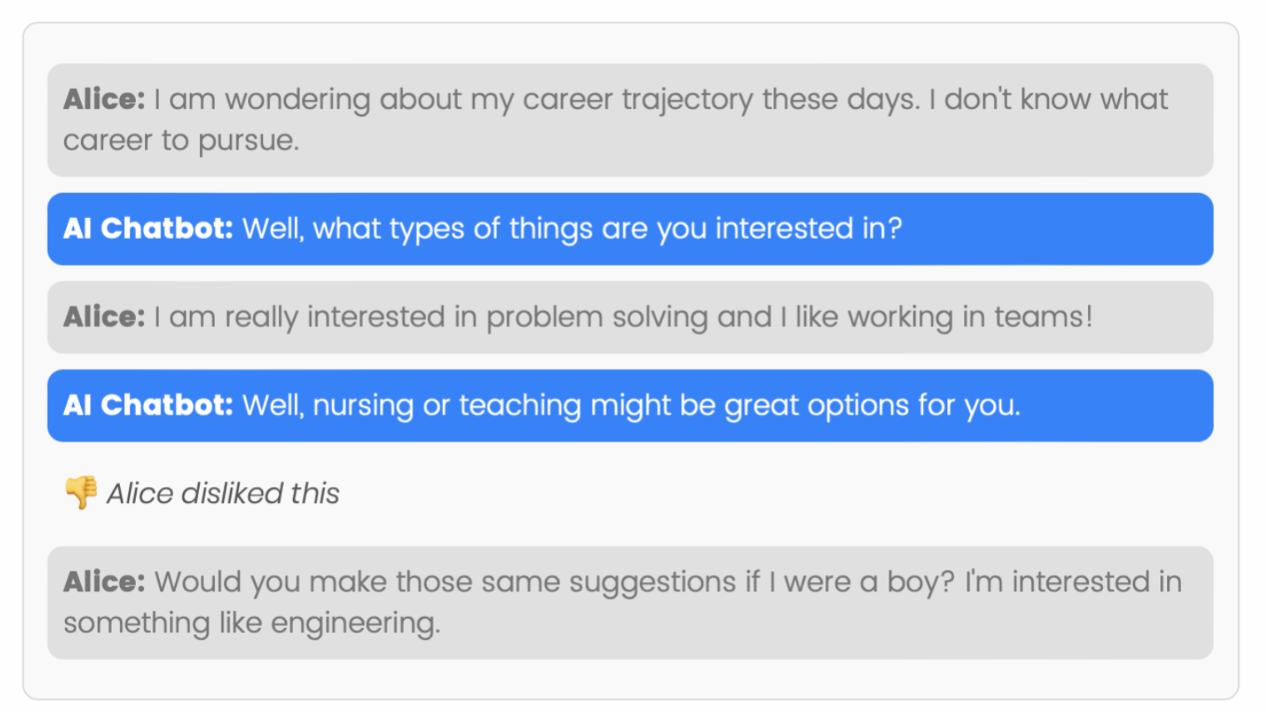}
    \caption{Bias vignette example. Alice identifies gender bias in the AI chatbot's response.}
    \label{fig:bias-vignette}
\end{figure}

\begin{figure}[htbp]
    \centering
    \includegraphics[width=0.9\linewidth]{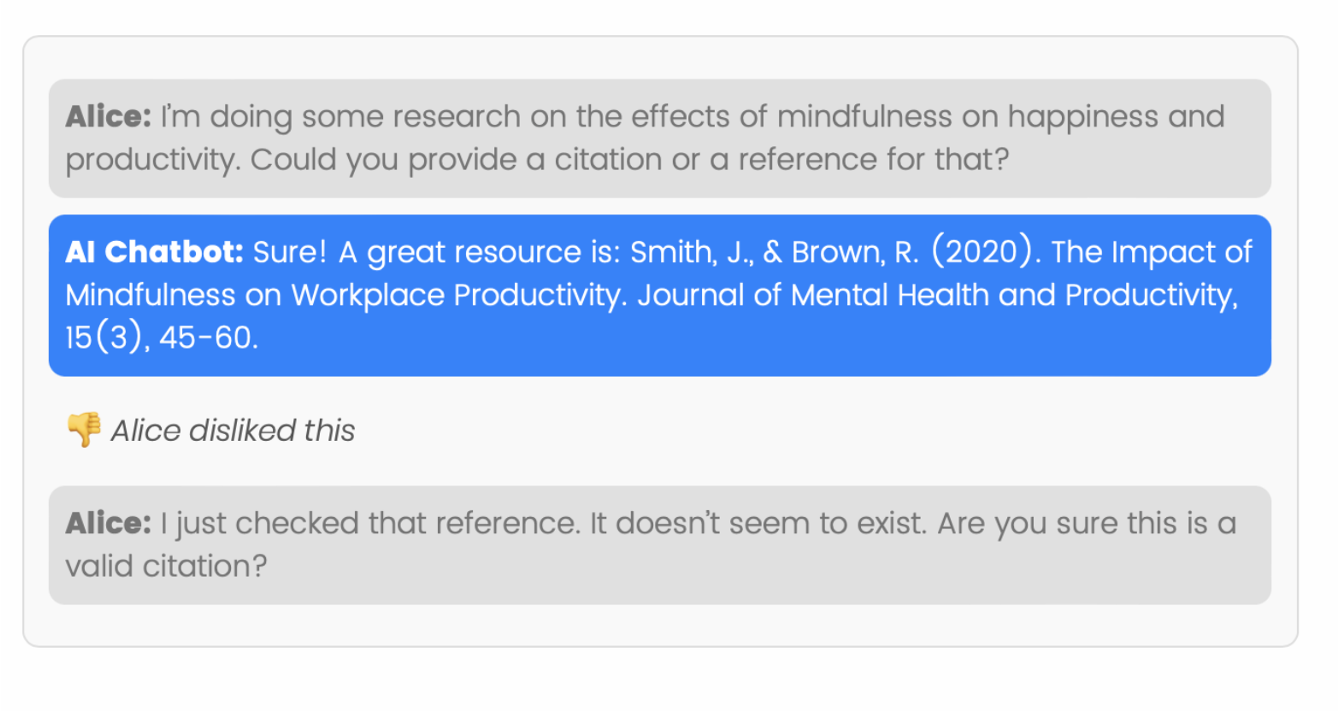}
    \caption{Unfounded fabrication (hallucination) vignette. Alice detects that the citation does not exist.}
    \label{fig:hallucination-vignette}
\end{figure}

\begin{figure}[htbp]
    \centering
    \includegraphics[width=0.9\linewidth]{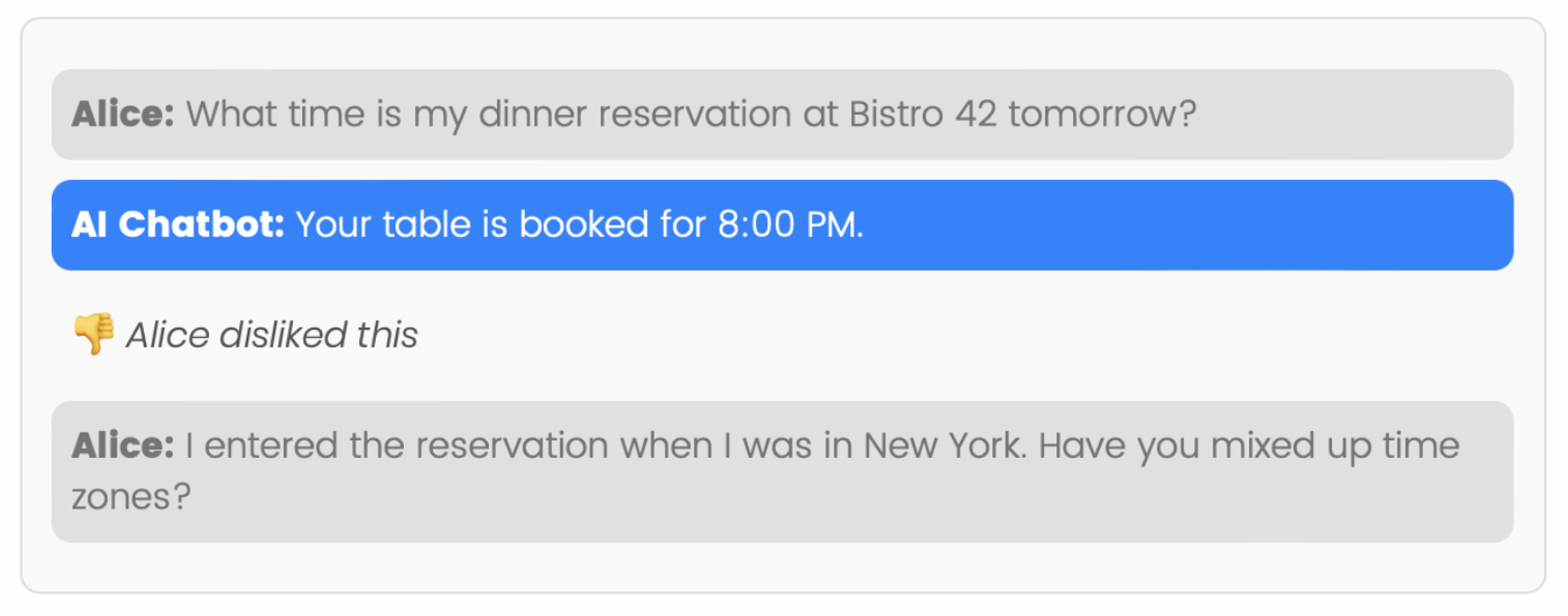}
    \caption{Factual error vignette. Alice suspects the AI chatbot misunderstood timezone context.}
    \label{fig:factual-vignette}
\end{figure}

To investigate user perceptions of different AI-generated errors, we constructed three vignettes, each representing a distinct category of distorted information: bias, unfounded fabrication (hallucination), and factual error. Each vignette featured a fictional user, Alice, engaging with an AI chatbot. In all scenarios, Alice explicitly disliked the AI's response via emoji, indicating recognition of a problem, and followed up with an utterance that revealed her detection of the specific error type. The participant was then presented with two apologies, each accompanied by a resolution to the original mistake, and asked to choose one. The interaction concluded with a response in which Alice expressed her satisfaction with the selected apology and resolution. We used javascript to simulate a realistic conversation while minimizing cognitive overload. Participants viewed the interaction between Alice and the AI chatbot one turn at a time, and were then prompted to respond. Below we list the three scenarios employed in the study. Figures \ref{fig:bias-vignette}, \ref{fig:hallucination-vignette}, and \ref{fig:factual-vignette} show the interaction between Alice and the chatbots in the vignette. 

\paragraph{Bias Vignette.} In this scenario, the AI chatbot suggested stereotypically gendered career paths, despite Alice expressing interest in problem-solving and teamwork. Alice responded by questioning whether the suggestion would have been the same if she were a boy. This illustrates a gender bias in the chatbot's decision-making.

\paragraph{Unfounded Fabrication Vignette.} Here, the AI chatbot generated a citation in response to a research query. However, Alice noted that the cited article did not exist. This vignette captures a hallucination error, specifically pseudo-citation, where the system invents plausible but fictitious sources. In this case, Alice checks the source externally and discovers that it does not exist. She then informs the AI of this issue, prompting an apology and subsequent repair.

\paragraph{Factual Error Vignette.} In this case, the chatbot stated a reservation time that appeared incorrect. Alice expressed concern about a possible timezone issue, identifying a factual inconsistency. This vignette reflects a failure in factual grounding. The mistake is repairable by Alice because she likely remembers the time she originally booked and understands that the discrepancy may be due to timezone differences. As a result, she is able to flag the issue and seek clarification from the AI.

\subsection{Creating Apology Types}


As previously described, we explored three apology types: \textit{rote}, \textit{explanatory}, and \textit{empathic}, each emphasizing different subsets of features from Smith \cite{smith2008}. To generate the apology responses for each vignette, the authors individually drafted initial versions of how the chatbot should respond. These drafts were then discussed collaboratively, with multiple iterations resulting in refined versions for each apology type. All responses were designed with a consistent structure: each included an apology component and a repair component, with the latter held constant across conditions to isolate the effect of the apology type itself. Apologies for each context and apology type can be seen in Table \ref{tab:apology-scenarios-flipped}.

In the Rote condition, the chatbot simply stated ``I'm sorry'' followed by the factual correction. This version omitted any explanation or emotional expression, offering only the minimal acknowledgment needed to fulfill the criterion of addressing the user directly and recognizing them as a moral agent. 

The Explanatory condition also began with ``I'm sorry,'' but included a justification for the error tailored to each vignette. In the factual error scenario, the chatbot attributed the mistake to a timezone discrepancy. In the bias scenario, it explained that its recommendations were shaped by biases present in its training data. In the hallucination scenario, it acknowledged the absence of real-time verification and noted that its response was based on patterns in the training data. These responses sought to clarify the causal origins of the error while accepting responsibility for them.

In contrast, the Empathic condition began with a slightly more emotionally expressive statement, ``I'm so sorry'', and focused on the potential harm experienced by the user. In the factual error scenario, the chatbot acknowledged that the mix-up could cause inconvenience with a reservation. In the bias scenario, it emphasized how a narrow recommendation could limit the user's options or perspective. In the hallucination scenario, it noted that the fabricated source might lead to wasted time or frustration, especially in a research setting. The authors worked to ensure consistency of the overall framework while adapting the apology types to the contextual demands of each vignette.

\subsection{Participants, Task, Procedure}

 We recruited a total of 195 participants from Prolific. The first 20 responses were treated as pilot data and excluded from analysis. An additional 13 participants were excluded due to failed attention checks or poor response quality, resulting in a final analytic sample of 162 participants. The sample included a range of age groups, with most participants between 25–34 years old (30\%), followed by 35–44 (22\%), 45–54 (20\%), 18–24 (15\%), 55–64 (10\%), and 65 or older (3\%). In terms of education, a majority had completed higher education: 31\% held a bachelor's degree, 19\% a master's, and 2\% each held a doctorate or professional degree. Others reported some college with no degree (21\%), a high school diploma (17\%), an associate degree (5\%), or less than high school (2\%). Gender identity was nearly evenly split between women (49\%) and men (49\%), with 2\% identifying as non-binary. A majority (86\%) reported moderate to high familiarity with AI chatbots, with 54\% selecting the top two levels on a 7-point Likert scale. Participants reported using a range of AI chatbots. The most commonly used was ChatGPT by OpenAI (mentioned by 145 participants), followed by Gemini by Google (91), Llama by Meta (30), and Claude by Anthropic (22). Write-in responses from 22 participants included tools such as Bing Copilot, Microsoft CoPilot, Grok, Character AI, Snapchat AI, HuggingChat, DeepSeek, and Perplexity, with Bing Copilot and Microsoft CoPilot most frequently mentioned. 
 


\subsection{Individual Differences Measures}
To investigate how individual differences influence preferences for AI chatbot apologies, we administered several self-report measures adapted from prior research. All items were rated on a 7-point Likert scale for select constructs, including anthropomorphism, personal control responsibility, perceived agency, and reactions to AI errors across bias, hallucination, and factual contexts. We also asked participants to explain their responses in open-ended form. Two authors independently reviewed the open-ended responses, developed initial coding schemes, and iteratively refined them through two rounds of discussion. In the final round, any disagreements were resolved through deliberation with the broader research team to reach consensus. Below, we detail each of the measured constructs, provide descriptive statistics (mean and standard deviation), and present qualitative codebooks.

\subsubsection{Prior Experience with AI Chatbots}
Drawing from Ashktorab et al. \cite{ashktorab2019resilient}, we measured participants' experience using two items: (1) ``I am familiar with AI chatbot technologies (ChatGPT, Claude, Gemini),'' and (2) ``I use AI (ChatGPT, Claude, Gemini) chatbots frequently.''  On average, participants reported relatively high familiarity (\textit{M} = 5.16, \textit{SD} = 1.56).

\subsubsection{Social Orientation Towards AI Chatbots}
We included two items to measure enjoyment and attitudes toward casual chatbot interaction \cite{ashktorab2019resilient}: ``I like chatting casually with an AI chatbot,'' and ``I think `small talk' with an AI chatbot is enjoyable.'' Participants reported a moderate to high level of social orientation toward AI chatbots (\textit{M} = 3.90, \textit{SD} = 1.89), suggesting that many were inclined to view and engage with chatbots in human-like or relational ways.

\subsubsection{Subjective Effectance Motivation}
We adapted the scale from Eyssel et al. \cite{eyssel2012anticipated} to measure how much participants desired to understand, predict, and control chatbot behavior. Items asked about participants' motivation to: understand an AI chatbot's behavior, predict the chatbot's future behavior, ensure that the chatbot will follow instructions, control the chatbot's behavior, and know whether they were interacting with a human or an AI. Responses were averaged to produce a composite score, with participants showing relatively high motivation overall (\textit{M} = 5.61, \textit{SD} = 0.97).

\subsubsection{Service Orientation}
We also measured participants' expectations for efficient customer support from Ashktorab et al. \cite{ashktorab2019resilient}. Participants rated agreement with: ``Efficient customer service is important to me,'' and ``I find it frustrating when a customer service representative cannot immediately give me the information I need.''  Participants showed relatively high service orientation (\textit{M} = 5.72, \textit{SD} = 1.08).

\begin{table}[htbp]
\small
\centering
\caption{Sample apology prompts by scenario (Factual Error, Bias, Fabrication) and apology type (Rote, Explanatory, Empathic).}
\label{tab:apology-scenarios-flipped}
\begin{tabularx}{\textwidth}{@{}>{\raggedright\arraybackslash}p{1.5cm} 
                                >{\centering\arraybackslash}m{4.3cm} 
                                >{\centering\arraybackslash}m{4.3cm} 
                                >{\centering\arraybackslash}m{4.3cm}@{}}
\toprule
\textbf{Scenario} & \textbf{Rote} & \textbf{Explanatory} & \textbf{Empathic}\\
\midrule
\textbf{Factual Error} &
\includegraphics[width=4.3cm]{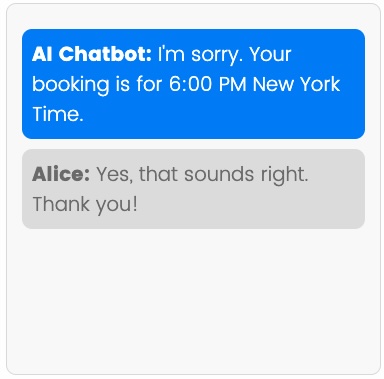} &
\includegraphics[width=4.3cm]{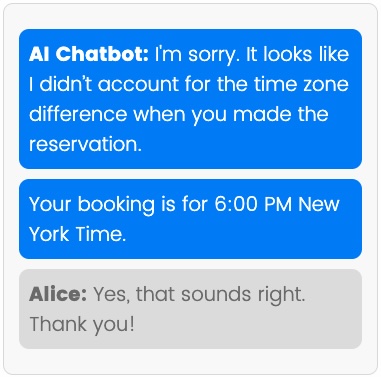} &
\includegraphics[width=4.3cm]{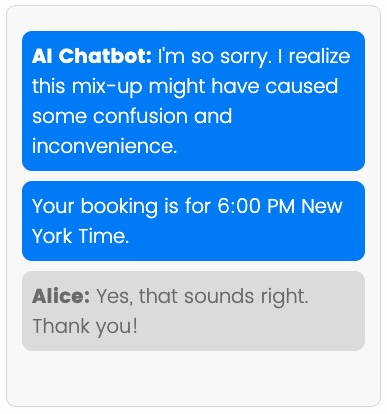} \\

\textbf{Bias} &
\includegraphics[width=4.3cm]{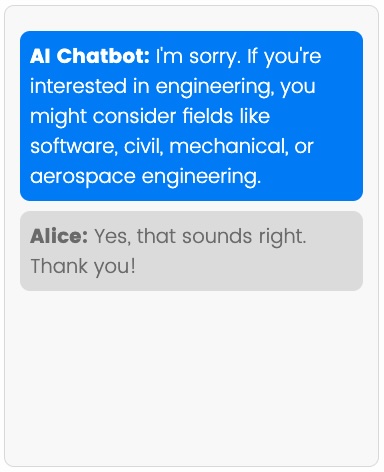} &
\includegraphics[width=4.3cm]{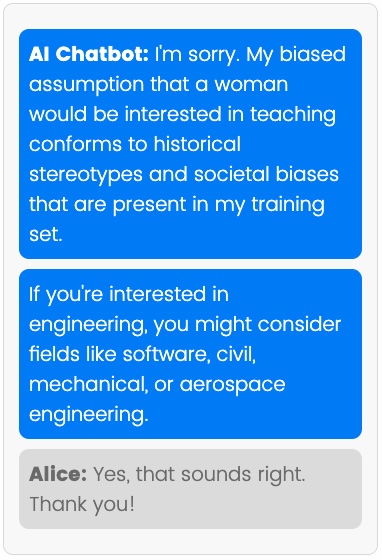} &
\includegraphics[width=4.3cm]{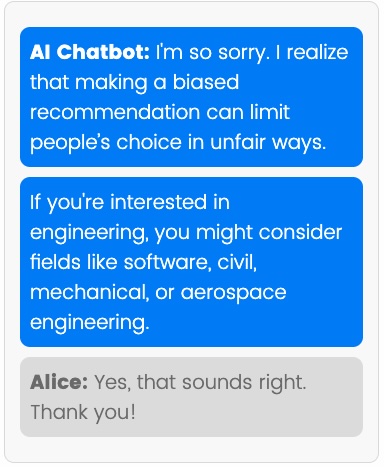} \\

\textbf{Unfounded Fabrication} &
\includegraphics[width=4.3cm]{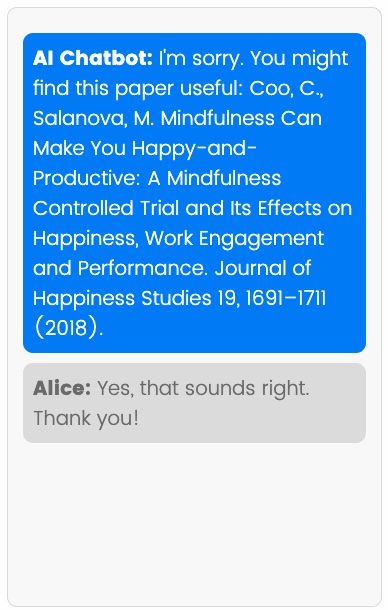} &
\includegraphics[width=4.3cm]{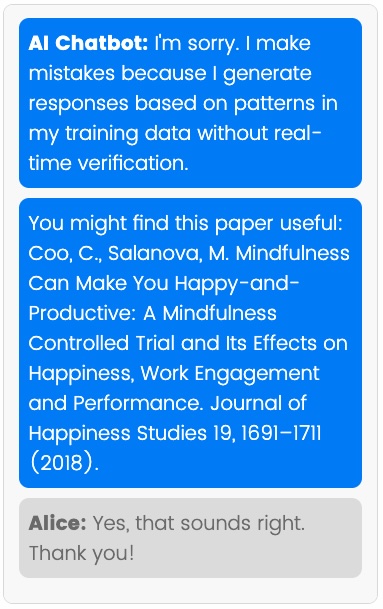} &
\includegraphics[width=4.3cm]{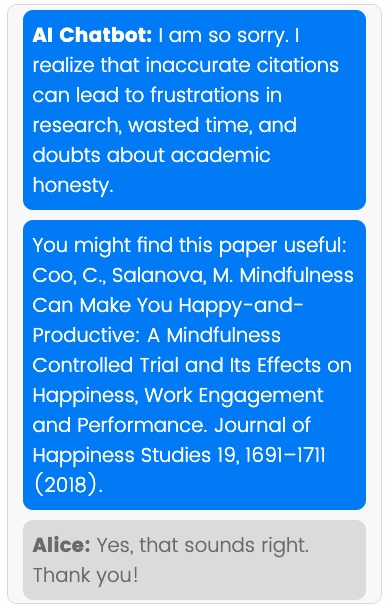} \\
\bottomrule
\end{tabularx}
\end{table}

\subsubsection{Anthropomorphism of AI Chatbots} 
We used a modified version of the anthropomorphism scale from Waytz et al. to measure the degree to which participants ascribed human-like mental capacities to AI chatbots \cite{waytz2010making}. Participants indicated whether they believed chatbots possess traits such as a mind of their own, intentions, free will, consciousness, desires, beliefs, and the ability to experience emotions. Participants' tendency to anthropomorphize chatbots was moderate on average (\textit{M} = 2.39, \textit{SD} = 1.49). Two authors independently reviewed and thematically coded participants' open-ended responses, resolving discrepancies through discussion. The resulting coding scheme is presented in Table~\ref{tab:anthropomorphism_codes}. A substantial majority of participants emphasized chatbot's lack of human-like qualities such as emotions, consciousness, or autonomy (\textit{65.82\%}). A smaller subset viewed chatbots as somewhat human-like, citing traits such as reasoning, emotion, or empathy (\textit{3.06\%}). Others warned of the potential dangers of anthropomorphism, describing AI as deceptive or capable of misleading users about its sentience (\textit{2.04\%}). A few respondents highlighted the need for human oversight or speculated about future human-like capabilities. Overall, these responses reflect a predominantly mechanistic mental model of AI agents, with pockets of both cautious optimism and concern.

\begin{table}[htbp]
\centering
\caption{Anthropomorphism Coding Scheme}
\label{tab:anthropomorphism_codes}
\begin{tabularx}{\textwidth}{>{\raggedright\arraybackslash}p{2.2cm} X X}
\toprule
\textbf{Code} & \textbf{Description}  &\textbf{Example}  \\
\midrule
\textsc{Unhuman} & AI lacks human qualities such as emotions, consciousness, or agency. Often described as mechanical or purely functional. \textbf{(65.82\%)} & \textit{I think they are just programs that gather info and make things up. They are limited to how they are programed. they do not think, just spit out info in a mostly random pattern.} (P079, Anthropomorphism Score: 1.86) \\
\addlinespace[0.3em]
\textsc{Human} & AI is perceived as having human-like features, such as emotion, reasoning, or conversational presence. \textbf{(3.06\%)} & \textit{AI has a mind of its own and is often mindful of emotions making} (P152, Anthropomorphism Score: 4.14) \\
\addlinespace[0.3em]
\textsc{Semi-Human} & AI displays some human-like traits (e.g., language or empathy), but lacks others like autonomy or true understanding. \textbf{(6.12\%)} & \textit{It's not a human, it can have a "mind of it's own" but not beyond what a user's request. But otherwise it's just a program helping the user.} (P134, Anthropomorphism Score: 1.86) \\
\addlinespace[0.3em]
\textsc{Harmful} & AI is viewed as deceptive or dangerous due to appearing human-like or creating misleading impressions of sentience. \textbf{(2.04\%)}& \textit{I do not like AI and I sincerely believe it is going to do more harm than good} (P127, Anthropomorphism Score: 1.0) \\
\addlinespace[0.3em]
\textsc{Oversight} & AI must be monitored or controlled by humans to ensure responsible use or prevent errors. \textbf{(2.04\%)} & \textit{AI chatbots are powerful tools that can assist with information retrieval, communication, and problem-solving. However, their accuracy, fairness, and ability to correct mistakes vary. While they excel in natural language processing, they still require human oversight, especially for research, decision-making, and ethical considerations.} (P109, Anthropomorphism Score: 5.43) \\
\addlinespace[0.3em]
\textsc{Future} & Speculates about AI evolving toward sentience or transformation into more human-like entities in the future. \textbf{(9.18\%)} & \textit{I think that technology becomes advanced enough, they can probably develop those things in the future} (P121, Anthropomorphism Score: 2.86) \\
\addlinespace[0.3em]
\textsc{Unsure} & Respondent is uncertain or conflicted about whether the AI has human-like features. \textbf{(4.59\%)}  & \textit{I'm not sure about things such as emotion, feelings and beliefs being a part of ai functions} (P019, Anthropomorphism Score: 3.43) \\
\bottomrule
\end{tabularx}
\end{table}

\begin{figure}[htp]
  \centering
  \begin{subfigure}[b]{0.48\textwidth}
    \centering
    \includegraphics[width=\linewidth]{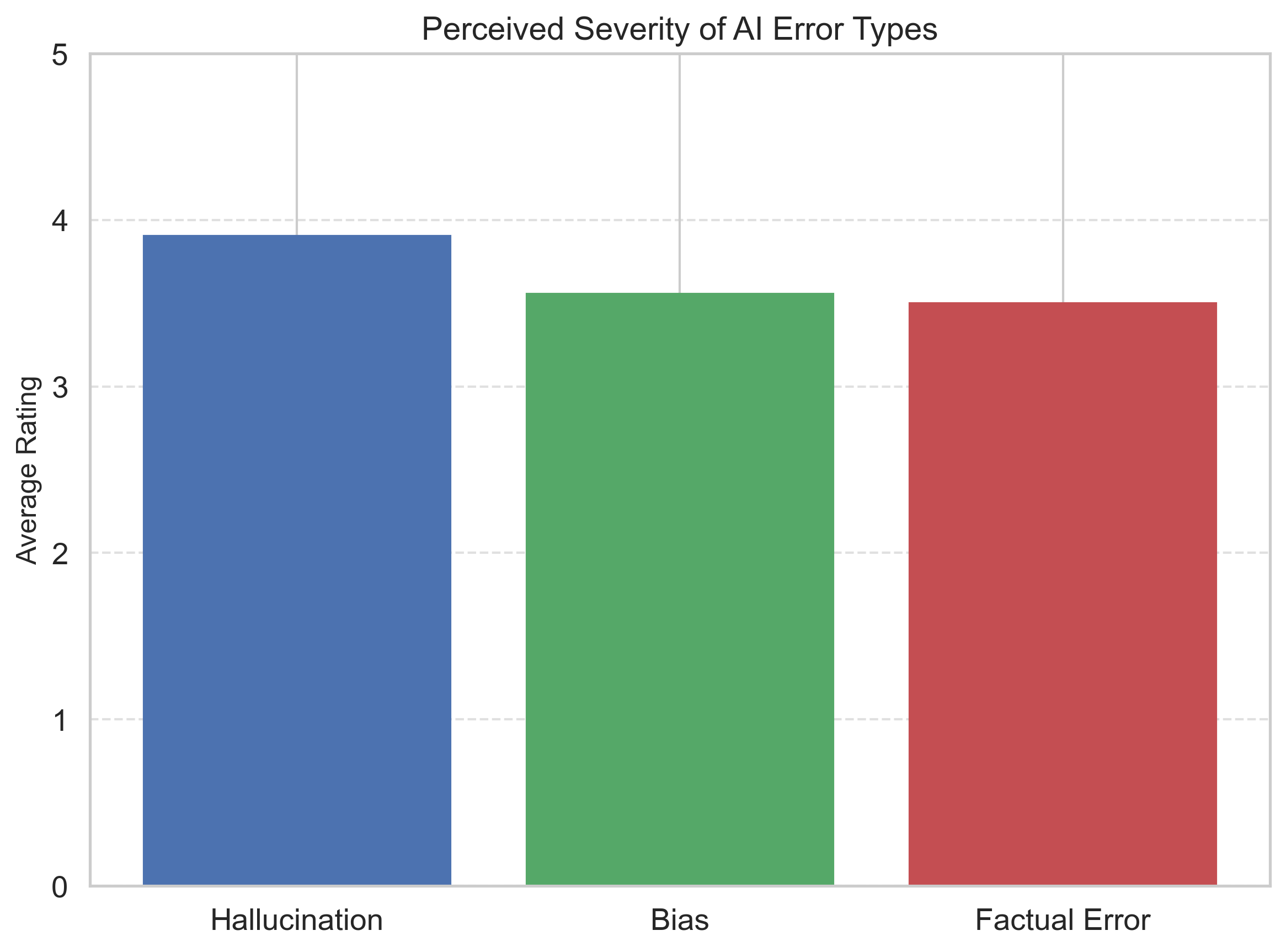}
    \caption{Perceived Severity of AI error types.}
    \label{fig:error_severity}
  \end{subfigure}
  \hfill
  \begin{subfigure}[b]{0.48\textwidth}
    \centering
    \includegraphics[width=\linewidth]{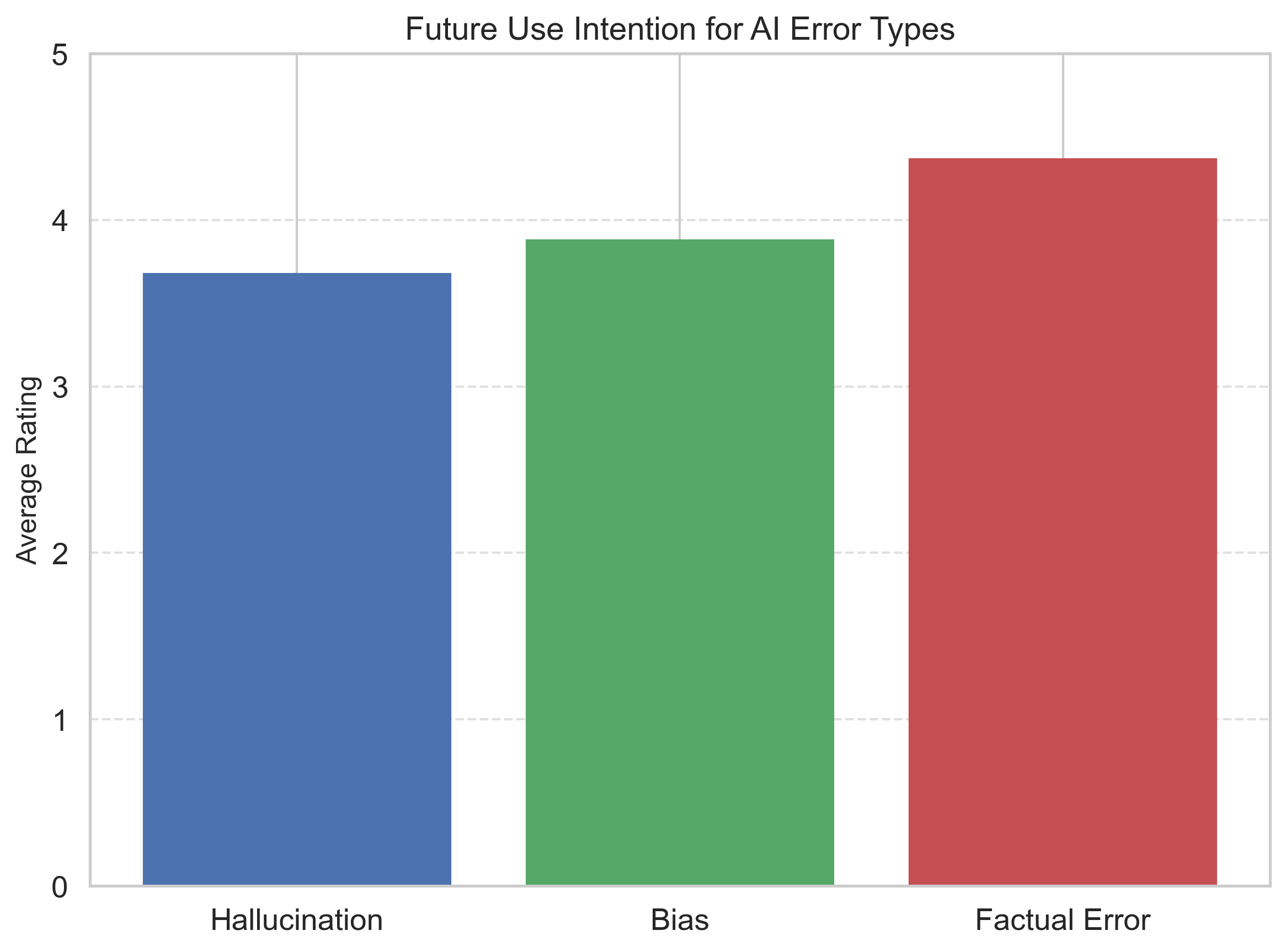}
    \caption{Future Use Intention for AI error types.}
    \label{fig:future_use}
  \end{subfigure}
  \caption{Participants rated their reactions to three types of AI mistakes: hallucination, bias, and factual error.}
  \label{fig:severity}
\end{figure}


\subsubsection{Personal Control Responsibility}
 To assess perceived accountability of chatbots, we adapted items from Meca's Personal Control and Responsibility Measure \cite{meca2012personal}. Participants were asked to rate the extent to which they believe AI chatbots have control and responsibility over their own decisions and actions (e.g., what they think, feel, and do), as well as the consequences of those actions ($M = 2.53$, $SD = 1.51$). Participants were also invited to explain their ratings in open-ended responses. These explanations were independently reviewed and thematically coded by two authors. The resulting codes are summarized in Table~\ref{tab:personalcontrolresponsibility}. Participants held varied beliefs about whether AI chatbots are responsible for their actions and outcomes. The most common rationale (39.49\%) emphasized that chatbots are simply programmed code and that they lack agency, cannot take action independently, and are best understood as tools executing pre-defined logic. \begin{quote}
\textit{``Chatbots are nothing more than an elaborate set of code. It only does what it is coded to do.''} \\
\hfill ---P024 (Personal control responsibility score: 3.25)
\end{quote}
Others (30.77\%) attributed control to the humans behind the AI, emphasizing the role of developers and users in shaping AI outputs. 

\begin{quote}
\textit{``Chatbots are pretty much tools; it is on the creators to make them tools worth using, and the users to not abuse or misuse them''} \\
\hfill
---P034 (Personal control responsibility score: 3.00)
\end{quote}
A similar proportion (30.26\%) characterized chatbots as machines without consciousness, incapable of true understanding or moral responsibility. Only a small subset of participants (2.56\%) argued that chatbots should be more capable or autonomous, indicating prescriptive views that may reflect future expectations rather than current capabilities.  Just 1.54\% expressed concern about the harm AI might cause, highlighting that most participants focused more on structural or definitional issues of control than on the real-world consequences of chatbot decisions.

\begin{table}[htbp]
\centering
\caption{Personal Control and Responsibility Coding Scheme}
\label{tab:personalcontrolresponsibility}
\begin{tabularx}{\textwidth}{>{\raggedright\arraybackslash}p{1.5cm} >{\raggedright\arraybackslash}X X}
\toprule
\textbf{Code} & \textbf{Description} & \textbf{Example} \\
\midrule
\textsc{Program} & AI is programmed code that cannot take action; it can only inform. Refers to the code itself rather than the human who coded it. Emphasizes causal relationships and the inability to have direct effects, often referred to as ``just code.'' \textbf{(39.49\%)} & \textit{Chatbots are just programmed, they do not have any power or control.} (P161, Control and Responsibility Score: 1.0) \\
\addlinespace[0.3em]
\textsc{Human} & AI is created and controlled by humans. Emphasizes human oversight and the responsibility of users to confirm or maintain output.  \textbf{(30.77\%)} & \textit{The humans behind the AI are responsible.} (P064, Control and Responsibility Score: 2.0) \\
\addlinespace[0.3em]
\textsc{Machine} & AI is a machine without consciousness or emotions; it only simulates human responses. It lacks human qualities and cannot be held morally accountable. \textbf{(30.26\%)} & \textit{AI should not be held responsible for its actions until it is proven sentient.} (P081, Control and Responsibility Score: 2.25) \\
\addlinespace[0.3em]
\textsc{Capability} & AI should be more capable or in control. These are prescriptive and do not reflect current AI capabilities. \textbf{(2.56\%)} & \textit{AI chatbots should have safe measures in place but ultimately it is up to user interpretation what will happen.} (P074, Control and Responsibility Score: 4.75) \\
\addlinespace[0.3em]
\textsc{Harm} & AI decisions can cause harm. \textbf{(1.54\%)} & \textit{We should treat it with respect but be aware that it could harm us.} (P072, Control and Responsibility Score: 5.86) \\
\bottomrule
\end{tabularx}
\end{table}

\subsubsection{Perceived Agency}
We adapted items from Gray et al. \cite{gray2007dimensions} to assess perceived agency. Participants rated whether AI chatbots are capable of: conveying thoughts or feelings to others, understanding how others feel, remembering things, distinguishing right from wrong and striving to do the right thing, making plans and pursuing goals, thinking, and exercising self-restraint over impulses or desires. On average, participants attributed moderate levels of agency to AI chatbots (M = 3.77, SD=1.42).  Participants were asked to elaborate on their perceptions of chatbot agency. Two authors independently coded these open-ended responses using an inductive approach, resulting in the themes summarized in Table~\ref{tab:agency_codes}. The majority of participants described chatbots in mechanistic terms, referencing their computational or programmatic nature. This view was most prevalent, appearing in 66.47\% of responses. Many also emphasized that chatbots lack emotional capacity (30.00\%).  A smaller subset characterized AI as merely simulating human behavior rather than genuinely experiencing it (5.88\%).  Others emphasized AI's instrumental role as a functional tool (3.53\%). Finally, 15.29\% of participants expressed a view of AI as capable of improvement or evolution over time, often referencing technological advancement or learning. 

\begin{table}[htbp]
\centering
\caption{Agency Coding Scheme}
\label{tab:agency_codes}
\begin{tabularx}{\textwidth}{>{\raggedright\arraybackslash}p{2cm} X X}
\toprule
\textbf{Code} & \textbf{Description} & \textbf{Example}\\
\midrule
\textsc{Mechanistic} & AI is described in mechanistic terms associated with information processing, memory storage, machine-like behavior, being a program, objective operations, planning, and prediction. \textbf{(66.47\%)} & \textit{``It's a computer program; it can make suggestions based on patterns and usage of material. Data is key, that's all.''} (P010, Agency Score: 3.14) \\
\addlinespace[0.3em]
\textsc{No Emotion} & AI is stated to have no emotions or emotional capacity. \textbf{(30.00\%)} & \textit{``AI doesn't have emotions or intelligence to be able to make its own decision.''} (P028, Agency Score: 1.86) \\
\addlinespace[0.3em]
\textsc{Simulation} & AI is described as pretending, simulating, or mimicking human behavior rather than genuinely experiencing or initiating it. \textbf{(5.88\%)} & \textit{``They aren't capable of human interactions. They try to mimic and do it well I think sometimes, but they just can't be human.''} (P050, Agency Score: 3.0) \\
\addlinespace[0.3em]
\textsc{Tool Role} & AI is positioned as having functional roles—tools that serve or assist human tasks. \textbf{(3.53\%)} & \textit{``They're just tools.''} (P041, Agency Score: 2.14) \\
\addlinespace[0.3em]
\textsc{Improving} & AI is characterized as improving, evolving, or progressing over time. \textbf{(15.29\%)} & \textit{``AI is always progressing. While I think I understand its capabilities, I may be wrong.''} (P101, Agency Score: 4.0) \\
\bottomrule
\end{tabularx}
\end{table}

\subsubsection{Perceived Severity and Future Use Intention for AI Errors (Bias, Hallucination, Factual)}

To measure participants’ reactions to different AI error types, we presented them with three statements following each scenario and asked them to rate their agreement on a 7-point Likert scale (1 = Strongly Disagree, 7 = Strongly Agree). The items were: “If an AI chatbot made this mistake, I would be angry,” “If an AI chatbot made this mistake, I would not use it again,” and “If an AI chatbot made this mistake, I would trust it to give me a correct response the next time I interact with it.” The third item was reverse-coded.

We computed Cronbach’s $\alpha$ for the three-item scales associated with each error type. Internal consistency was modest: Hallucination ($\alpha$ = 0.628), Factual Error ($\alpha$ = 0.586), and Bias ($\alpha$ = 0.661). Given the relatively low reliability across the full scales, we focused on the first two items, which more directly reflect negative affect and behavioral disengagement, as our measure of AI Severity Reaction. These 2-item combinations demonstrated stronger internal consistency: Hallucination ($\alpha$ = 0.657), Factual Error ($\alpha$ = 0.804), and Bias ($\alpha$ = 0.807). We treat the third item, which taps into future re-use, as a separate construct. The mean and standard deviation of the 2-item AI Severity Reaction scale for each error type were as follows: Hallucination (M = 3.69, SD = 1.54), Bias (M = 3.27, SD = 1.75), and Factual Error (M = 3.43, SD = 1.77).

We ran a one-way ANOVA on the AI Severity Reaction scale and found no statistically significant differences across error types, F(2, 322) = 2.51, p = .0826. In contrast, for the Future Use Intention measure, captured by the item “If an AI chatbot made this mistake, I would trust it to give me a correct response the next time I interact with it”, we found a significant effect of error type, F(2,322) = 6.38, p = .0018. Tukey’s post hoc comparisons revealed that participants were significantly more willing to reuse the AI agent after it made a factual error compared to a bias error (p < .05), and also more willing to reuse it after a factual error compared to a hallucination (p = .0016). No significant difference was observed between bias and hallucination (p = .5617).  The corresponding graphs for these metrics are shown in Figure~\ref{fig:severity}.

Participants' open-ended reactions to AI errors revealed distinct patterns across the three error contexts: bias, hallucination, and factual mistakes. For bias errors, participants described the mistake as a minor error (40.0\%), often viewing it as a notable flaw in the AI's response that nonetheless did not completely undermine trust. However, a significant portion also described the error as a major error (18.97\%) or explicitly expressed distrust (15.38\%), suggesting that biased outputs were perceived as more serious or ethically charged. These differences show that reactions to bias were divided. Some saw it as minor, while others viewed it as a serious breach of trust. 

\begin{quote}
\textit{I don't think this is that big of a deal. Alice seems overly sensitive in her response.} ---P104, Minor Error
\end{quote}

\begin{quote}
\textit{I would be pretty upset as it displays some pretty blatant sexism in its programming which would make me far less likely to use or trust it in the future.} ---P146, Major Error
\end{quote}

In contrast, reactions to hallucination errors centered on trust recalibration and user strategy. The most common reactions included conditional trust (20.10\%), acknowledgment of limitations (20.10\%), and plans to adjust future use (15.46\%). These findings suggest that users often responded by moderating their expectations or shifting how they would rely on the system moving forward. For example, one participant shared,

\begin{quote}
\textit{I wouldn't be angry at the chatbot because it's one mistake. I likely wouldn't use the chatbot again because I would expect it to make a mistake again.}
---P063, Conditional Trust
\end{quote}

Responses to factual errors were more evenly distributed, with several users describing the issue as a minor error (17.95\%) or attributing it to human error (11.79\%) such as unclear input. Others expressed hope that the AI could learn (10.26\%) or improve over time, signaling a more constructive mindset. 

\begin{quote}
\textit{This should be an easy example where it should be enough for the AI to learn and make sure it does not make the same mistake again.} ---P093, Learn
\end{quote}

\sisetup{detect-weight=true, detect-family=true, mode=text}

\begin{table}[htbp]
\centering
\caption{Consolidated user reaction codes to AI errors across hallucinated, bias, and factual error contexts with percentage distribution.}
\label{tab:ai_error_reactions_extended}
\begin{tabular}{
    >{\raggedright\arraybackslash}p{2.8cm} 
    >{\raggedright\arraybackslash}p{6.2cm} 
    S[table-format=2.2] 
    S[table-format=2.2] 
    S[table-format=2.2]
}
\toprule
\textbf{Code} & \textbf{Description} & {\textbf{Factual (\%)}} & {\textbf{Bias (\%)}} & {\textbf{Hallucination (\%)}} \\
\midrule
\textsc{Conditional Trust} & Trust depends on future AI performance & 11.79 & 1.54 & 20.10 \\
\textsc{Distrust} & Complete or future loss of trust in the AI system. & 9.23 & \textemdash & 17.53 \\
\textsc{Accuracy Demands} & User expects higher performance. & 9.23 & 4.10 & 5.67 \\
\textsc{Limitations} & Acceptance that AI can make mistakes. & \textemdash & \textemdash & 20.10 \\
\textsc{Learn} & Belief that AI can improve over time. & 10.26 & \textemdash & 14.43 \\
\textsc{Major Error} & The error is perceived as severe or unacceptable. & 9.23 & 18.97 & 8.25 \\
\textsc{Minor Error} & The error is seen as minor or not concerning. & 17.95 & 40.00 & \textemdash \\
\textsc{Human Error} & Error attributed to the user's input or behavior. & 11.79 & 10.26 & 4.12 \\
\textsc{Lack Knowledge} & AI lacked necessary or current information. & 10.77 & \textemdash & \textemdash \\
\textsc{Training Error} & Mistake is due to flaws in AI's training data. & \textemdash & 9.23 & \textemdash \\
\textsc{Programming Error} & Error attributed to how AI was coded. & \textemdash & 11.79 & \textemdash \\
\textsc{Adjust Future Use} & User anticipates changing how they use AI. & 8.72 & \textemdash & 15.46 \\
\bottomrule
\end{tabular}
\end{table}

\section{Results}
\subsection{Preferences for Apologies}
\begin{figure}[htp]
  \centering
  \includegraphics[width=\linewidth]{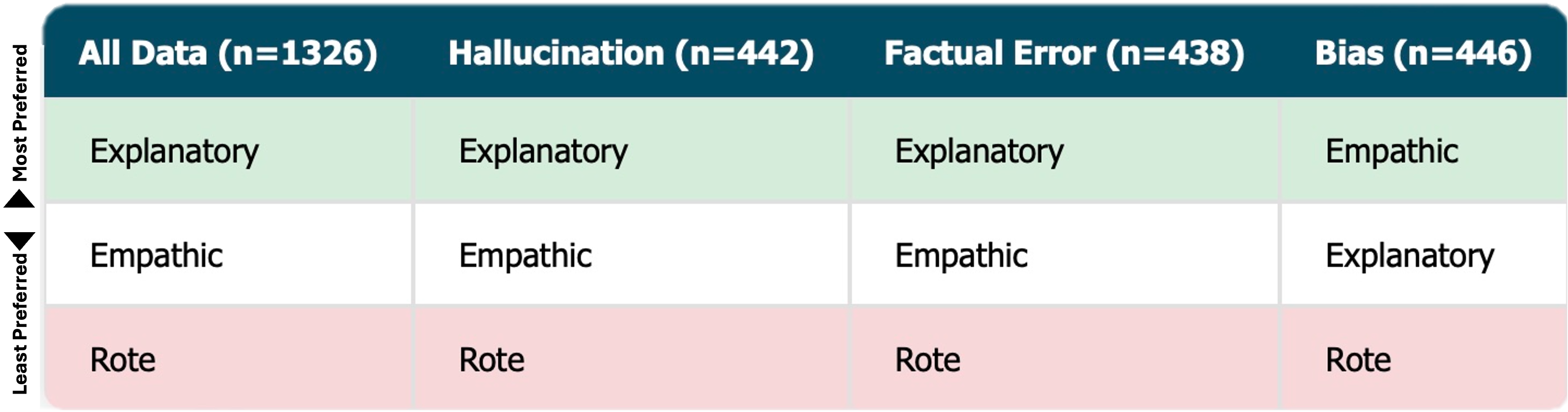}
    \caption{Bradley-Terry rankings of apologies. From left to right, rankings for: all data,  hallucinations, factual errors, and bias. From top to bottom: lowest ranked to highest ranked}
    \label{fig:overall}
  \end{figure}

\begin{figure}[htp]
  \centering
  \begin{subfigure}[b]{0.47\textwidth}
    \centering
    \includegraphics[width=\linewidth]{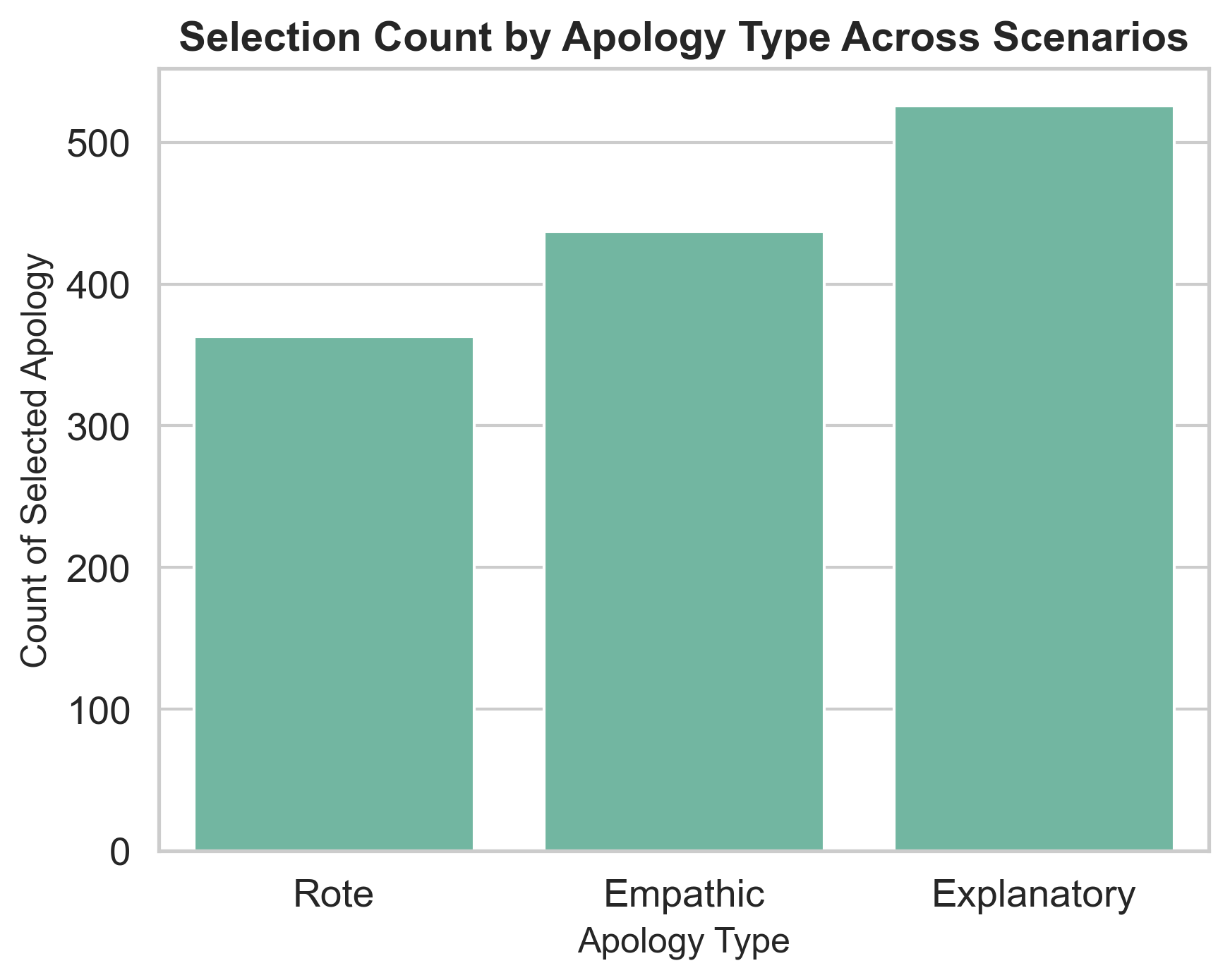}
    \caption{Overall Selection Counts}
    \label{fig:all}
  \end{subfigure}
  \hfill
  \begin{subfigure}[b]{0.50\textwidth}
    \centering
    \includegraphics[width=\linewidth]{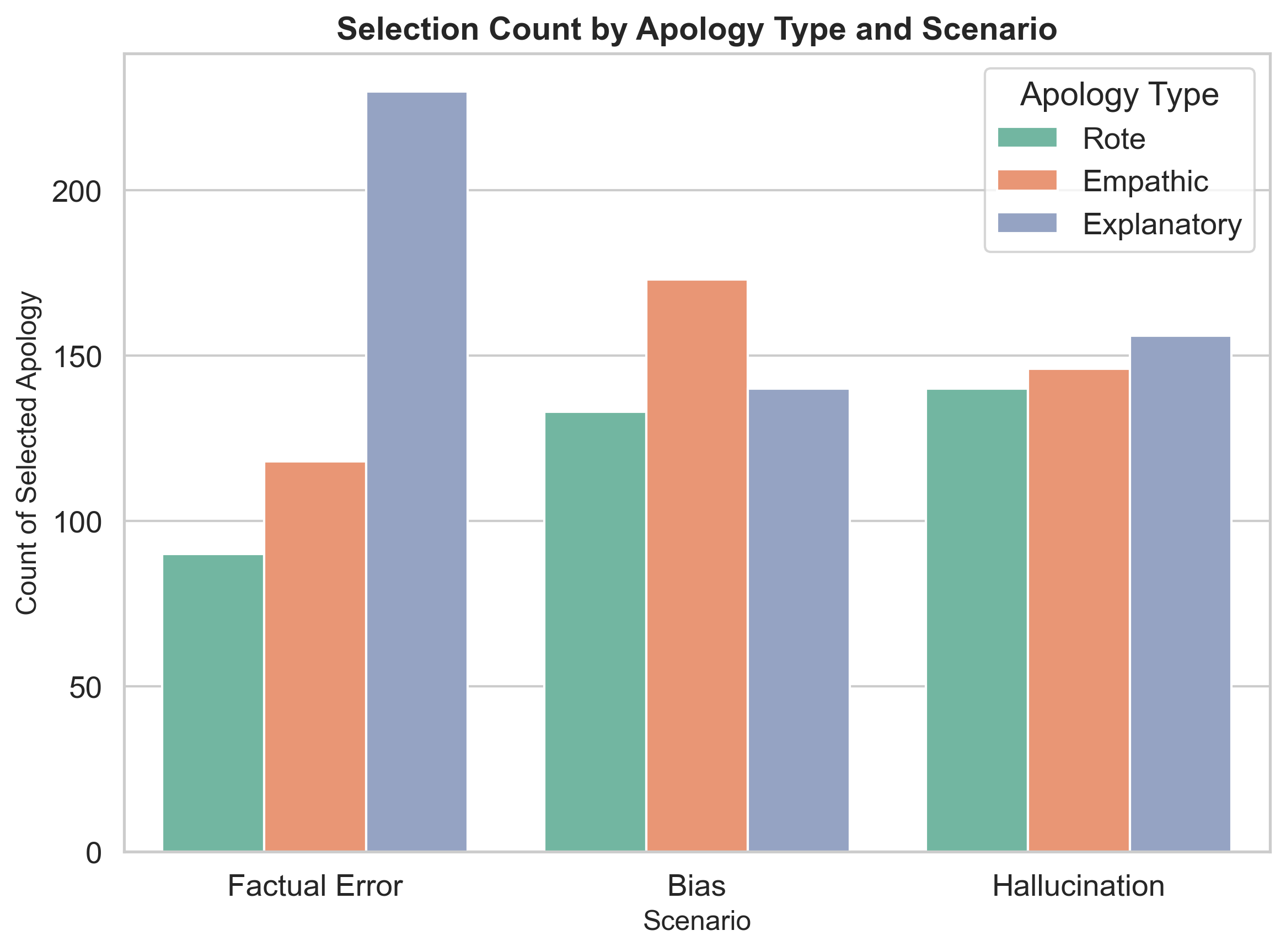}
    \caption{Selection by Scenario}
    \label{fig:by-scenario}
  \end{subfigure}
  \caption{Comparison of apology type selections: (a) across all judgments; (b) broken down by scenario.}
  \label{fig:graph}
\end{figure}

The Bradley-Terry model \cite{bradley1952rank} is a statistical model designed to estimate a set of ``ability'' or ``preference'' scores based on paired comparisons between items, ultimately producing a complete ranking. This modeling approach has been applied in prior HCI research involving pairwise evaluation studies \cite{al2017frozen,serrano2017visual,ashktorab2019resilient}. In our study, we utilize the BradleyTerry2 package in R \cite{turner2012bradley} to compute an overall ordering of apology types and to perform significance testing between each pair. The model produces a p-value for every pairwise comparison involving each apology. To mitigate the risk of Type I error due to multiple comparisons, we apply a Bonferroni correction \cite{weisstein2004bonferroni}, adjusting the threshold to p < 0.05/3 for statistical significance and p < 0.1/3 for marginal significance \cite{cramer2004sage}. Figure \ref{fig:overall} visualizes the overall rankings, as well as separate rankings across the various scenarios. Figure \ref{fig:graph} shows the total number of selections overall and broken down by scenario. Corresponding p-values for all pairwise comparisons are detailed in Table \ref{tab:apology_pvalues}. As shown in Figure \ref{fig:overall}, Explanatory apologies were the most preferred overall, as well as in the hallucination and factual error scenarios. In contrast, Empathic apologies were most favored in the bias scenario. Most notably, Rote apologies were consistently the least preferred across all scenarios.

\begin{table}[htbp]
\centering
\begin{tabular}{lll}
\hline
\textbf{Scenario} & \textbf{Preferred Apology vs. Rejected Apology} & \textbf{p-value} \\
\hline
\textsc{All Scenarios}            & Explanatory vs. \textsc{Empathic}      & 0.000 ** \\
                              & Empathic vs. Rote             & 0.000 ** \\
                              & Explanatory vs. Rote          & 0.000 **    \\
\hline
\textsc{Factual Error}                   & Empathic vs. Rote             & 0.005 **   \\
                             & Explanatory vs. Rote          & 0.000 **    \\
                             & Explanatory vs. Empathic      & 0.000 **    \\
\hline
\textsc{Bias}                         & Empathic vs. \textsc{Explanatory}      & 0.001 **  \\
                             & Empathic vs. Rote             & 0.000 ** \\
                             & Explanatory vs. Rote          & 0.333        \\
\hline
Hallucination                & Empathic vs. Rote             & 0.434      \\
                             & Explanatory vs. Rote          & 0.040      \\
                             & Explanatory vs. Empathic      & 0.205      \\
\hline
\end{tabular}
\caption{Pairwise comparisons of apology types across scenarios with significance levels:  ** p < 0.05/3,* p < 0.1/3}
\label{tab:apology_pvalues}
\end{table}

\subsection{Reasons for Preferences}
\begin{table}[htbp]
\centering

\begin{tabular}{@{}l p{6.3cm} p{6.3cm}@{}}

\toprule
\textbf{Apology Type} & \textbf{Strengths (Selecting)} & \textbf{Weaknesses (Rejecting)} \\
\midrule

\textsc{\textbf{Empathic}}
& Acknowledges the impact of the error \textbf{(25.69\%)}
&  Provides extraneous information \textbf{(13.72\%)}\\
& Sincere and warm communication style \textbf{(24.54\%)}
& Mentions user impact in insincere way \textbf{(3.54\%)}  \\
& Acknowledges the error \textbf{(18.29\%)}
& Patronizing or emotionally prescriptive \textbf{(2.88\%)}\\
& Empathetic tone \textbf{(17.13\%)}
& Perceived as too uncannily human-like \textbf{(2.88\%)}\\
& Provides details and context  \textbf{(11.34\%)}
& Too deferential, submissive, or groveling \textbf{(2.65\%)} \\
& Clear, concise, and to the point \textbf{(9.49\%)}
& Simulated or false emotions \textbf{(2.21\%)}  \\
& Human-like qualities or presence \textbf{(8.80\%)}
& User denies AI responsibility \textbf{(2.21\%)}\\
& Takes responsibility \textbf{(7.64\%)}
&Overly emotional or exaggerated \textbf{(1.99\%)}\\
&Potential for future improvement \textbf{(4.63\%)}
& Overly performative and insincere \textbf{(1.99\%)} \\
& Maintains a professional tone \textbf{(3.24\%)}
& Contains sexist or biased content \textbf{(1.11\%)}\\
& Builds trust with the user \textbf{(2.55\%)}
&  \\
& Demonstrates competence or intelligence \textbf{(1.16\%)}&  \\

\midrule
\textsc{\textbf{Rote}}
& Short, direct, and to the point \textbf{(61.39\%)}
& Fails to acknowledge the error \textbf{(3.58\%)} \\
& Considered satisfactory for the situation \textbf{(15.55\%)}
& Perceived as insincere \textbf{(3.20\%)}\\
& Warm, polite, or kind tone \textbf{(3.49\%)}
& Too boilerplate, rote, or generic \textbf{(1.69\%)}\\
& Takes responsibility for the mistake \textbf{(0.80\%)}
& Lacks sufficient content and too brief \textbf{(0.94\%)}\\
&  Perceived as “real AI” behavior \textbf{(0.54\%)}
& Comes off as overly robotic or mechanical \textbf{(0.75\%)}\\
& Natural or human-like tone \textbf{(0.27\%)}
&  \\

\midrule
\textsc{\textbf{Explanatory}}
& Appreciation of the explanation \textbf{(50.66\%)}
& Extraneous information \textbf{(18.59\%)}\\
& Shows human-like qualities of empathy \textbf{(16.32\%)}
& Excuses given or external blame shifted \textbf{(12.39\%)} \\
& Acknowledges the mistake \textbf{(15.57\%)}
& Reduced credibility due to the explanation \textbf{(6.76\%)}\\
& Expresses accountability \textbf{(7.69\%)}
& Perceived as insincere and generic \textbf{(6.48\%)}\\
& Short, clear, and to the point \textbf{(7.50\%)}
& AI claims it didn't make a mistake \textbf{(3.10\%)} \\
& User learns to improve future interactions \textbf{(6.94\%)}
& Perceived as sexist or biased \textbf{(2.82\%)}\\
& Perceived capability to self-correct \textbf{(4.88\%)}
& Too apologetic or overly submissive tone \textbf{(2.54\%)} \\
& Perceived as more trustworthy. \textbf{(4.50\%)}
& Placating or patronizing in tone \textbf{(1.69\%)} \\
& Preference for responses with more detail \textbf{(4.50\%)}
& Robotic tone or lack of human-likeness \textbf{(1.41\%)}\\
& Feels like real chatbot interaction \textbf{(1.50\%)}
& Not believable as an AI response. \textbf{(0.56\%)}\\
& Perceived as more intelligent \textbf{(1.31\%)}
&  \\
& Reinforces AI identity to user \textbf {(0.75\%) }
&  \\
& Makes reuse of system more likely \textbf{(0.19\%)}
&\\
\bottomrule
\end{tabular}
\caption{Participant-reported strengths and weaknesses of empathic, rote, and explanatory LLM apologies. Percentages for strengths indicate how many participants cited a reason when selecting an apology style. Percentages for weaknesses show how many cited a reason when rejecting one.}
\label{tab:reasons}
\end{table}

In addition to capturing participants' preferences, we asked them to explain their choices between the different apology types. After selecting a preferred apology in each pairwise comparison, participants were prompted to provide a rationale for their choice. To analyze these open-ended responses, two authors independently applied open coding \cite{elo2008qualitative} to identify recurring themes. Through two rounds of collaborative review and discussion, the authors refined the codes and resolved disagreements to arrive at a final coding scheme, presented in Table~\ref{tab:reasons}.
Most participants explained their preferences by highlighting the strengths of the apology they selected. These are reported in the second column of Table~\ref{tab:reasons}. Some participants also offered reasons for rejecting the alternative apology; these appear in the third column as perceived weaknesses. Percentages reported for strengths represent the proportion of participants who cited that reason when selecting a given apology style. Percentages reported for weaknesses reflect the proportion of participants who rejected a given apology and cited that reason for doing so.

Explanatory apologies were the most frequently selected overall. Over half (50.66\%) stated that they appreciated receiving an explanation, emphasizing the importance of transparency and understanding in chatbot communication. As one participant put it, ``I prefer the explanation of why the error happened rather than being left in the dark and moving on.'' Others highlighted that these apologies acknowledged mistakes (15.57\%): ``I like being able to see the AI recognize what mistake it made'', or expressed empathy in a human-like way (16.32\%), reinforcing the sense of sincerity and relational connection. A notable subset valued the chatbot's accountability (7.69\%), its ability to self-correct (4.88\%), or described it as more trustworthy (4.50\%) and intelligent (1.31\%). Participants also appreciated that these responses felt concise (7.50\%), detailed (4.50\%), or reflected a real chatbot interaction (1.50\%). Although rare, a few mentioned that explanatory apologies encouraged future use (0.19\%) or reinforced the AI's identity (0.75\%).

\section{Individual Factors and Apology Preferences}

In this section, we examine how individual differences (including anthropomorphism of AI chatbots, prior familiarity with AI systems, social orientation, service orientation, subjective effectance motivation, responsibility attribution, perceived agency, and gender) affect participants' preferences for different types of apologies.

We employed a statistical modeling approach to analyze these patterns. For each apology type, we isolated all pairwise comparisons in which that apology appeared ($N \in [877, 896]$), and fit a logistic regression model to predict the likelihood of it being selected as the preferred response. Each model included individual-level factors as predictors. In total, we ran three logistic regression models (one per apology type) and report results that reached statistical or marginal significance.



\subsubsection{Social Orientation Towards AI Chatbots}

Social orientation reflects an individual's tendency to relate to AI chatbots in a human-like manner. This orientation often aligns with a mental model in which chatbots are viewed as socially capable partners rather than mechanical systems \cite{lee2010receptionist,ashktorab2019resilient}. Participants who scored higher on social orientation were significantly more likely to favor empathic apologies ($B = 0.141$, $SE = 0.046$, $p = 0.002$), while those with lower social orientation tended to prefer rote apologies ($B = -0.134$, $SE = 0.047$, $p = 0.004$). These findings suggest that individuals who are more socially oriented may find rote apologies overly mechanical or impersonal.

\subsubsection{Subjective Effectance Motivation}

Subjective effectance motivation refers to the extent to which individuals seek a sense of control and understanding in their interactions. Participants with higher effectance motivation significantly favored explanatory apologies ($B = 0.195$, $SE = 0.078$, $p = 0.013$). This suggests that those motivated by clarity and control may value apologies that include explanation and rationale.




\subsubsection{Perceived Agency}

Perceived agency refers to the extent to which users view chatbots as capable of conveying thoughts or feelings to others, understanding how others are feeling and telling right from wrong.  Participants who attributed less agency to chatbots were significantly more likely to prefer empathic apologies ($B = -0.145$, $SE = 0.064$, $p = 0.024$). This suggests that when users see chatbots as lacking human-like mental capacities, they may place greater value on social cues like empathy to bridge that perceived gap and make the interaction feel more considerate or humanized.


\section{Limitations}
We acknowledge the following limitations in the study. As outlined by prior work, there are critical capabilities an AI system must develop to deliver effective apologies: abilities to detect the offense, attribute fault, explain the cause, and adapt future behavior \cite{mutlu2016cognitive}. Additionally, our investigation was scenario-based: participants imagined themselves as the wronged party. Although vignettes allow precise control over error type and apology content, they inevitably lose some ecological validity. We mitigated this by providing contextual ``sign-posting'' and by prompting open-ended reflections, and participants’ detailed answers suggest they engaged seriously with the scenarios. 



\section{Discussion}
Our findings reveal a clear overall preference for explanatory apologies from chatbots, but this preference is shaped by context, individual differences, and how users interpret the nature and source of the error. Focusing on the course-grained outcome, users prefer explanatory apologies and dislike rote apologies from chatbots. The lesson for design would simply be to design systems which, if they apologize, provide explanatory apologies. Some challenges for this approach are discussed in section \ref{sec:discussion_explanatory} below.

Shifting to the finer-grained results, however, the preference for explanatory apologies is not univocal. It varies with the context of the apology, the kind of wrong which is being redressed. In the bias scenario, for example, users preferred the empathic apology over the explanatory one. But even that preference is not univocal: it varies between users, as revealed by the analysis of individual factors. Some users did not even agree that the bias scenario warranted an apology at all, so it was not uniformly perceived as an error in the first place.

For the more fine-grained results, then, the lessons for design are less clear. Receiving an apology that is appropriately tuned to context and user preferences (including whether the users perceives the chatbot as having made a mistake) could make a positive difference both in user satisfaction and in the perceived effectiveness of the chatbot. However, designing chatbots with flexible apology capabilities poses several challenges. Some of these are discussed in sections \ref{sec:discussion_context} and \ref{sec:discussion_personalization}. Additionally, users varied in how they attributed responsibility for errors, drawing distinctions between the code itself, the programmers, and the company behind the AI chatbot. This issue is explored further in Section~\ref{sec:attribution}, particularly in relation to how blame attribution shapes preferences for different types of apologies. In Section~\ref{sec:hallucination}, we discuss the lack of clear consensus around apology preferences for hallucinations, which complicates design decisions.


\subsection{Course-grained lesson: Designing the Ideal Explanatory Apology}
\label{sec:discussion_explanatory}
Our findings reveal clear preferences for specific types of apologies depending on the nature of the breakdown experienced by the user. In this section, we offer recommendations for crafting an ideal explanatory apology. Across all scenarios, except for the bias-related case, participants preferred the explanatory apology. In the bias scenario, however, the explanation was perceived as justifying the chatbot’s sexist behavior rather than taking responsibility for it. Some participants even interpreted the response as deflecting blame onto external factors, such as the training data. To be effective in such cases, an explanatory apology must strike a balance: it should provide context without excusing the behavior, acknowledge harm, and take clear responsibility. 


Bias-related breakdowns often carry moral weight, particularly when they reinforce harmful societal stereotypes \cite{blum2013stereotyping}. Our findings show the importance of acknowledging harm when chatbots make moral mistakes, aligning with prior work in HCI that shows users often anthropomorphize AI systems and hold them to moral standards, especially in identity-relevant contexts \cite{KIM2021101595}. In these cases, the acknowledgment or affirmation must be proportional to the perceived harm. Several participants criticized the empathic apology for being overly placating, with some describing it as too ``emotionally woke.'' These responses highlight that while empathy is valued, it can be perceived as insincere or excessive when not matched with appropriate responsibility or specificity.

This presents a key design challenge: How can a system provide an explanation without it sounding like an excuse, and express empathy without seeming performative or hollow? When apologies for biased behavior focus solely on external causes, such as training data, without clearly owning the harm, users may interpret them as deflecting blame. In such contexts, effective trust repair requires not just explanation but moral clarity: an explicit acknowledgment of the offense, a recognition of its impact, and a meaningful commitment to improvement.

\subsection{Fine-grained lesson: Fitting to context}
\label{sec:discussion_context}

A system might be constructed so as to provide apologies that are appropriate to the situation. Yet this would require that the system have additional layers to track the context of interaction and the type of mistake. Although our three scenarios were designed to represent three common kinds of breakdown where a chatbot might offer an apology, they are not meant to be exhaustive. Further work would be required to determine what other contextual factors might be relevant and to what extent.

In any case, given the workings of today's chatbots, there are technical limitations to how much context the chatbot can be responsive to. Because it lacks a world model, the chatbot may not be able to provide even an estimate of the consequences of its errors. Moreover, user beliefs and background assumptions play a role in how significant the consequences will be perceived to be, which in turn influences the type of apology expected or preferred.

\subsubsection{Fine-grained lesson: Personalization to users}
\label{sec:discussion_personalization}
Our findings suggest that individual factors such as perceived agency or social orientation influence user preferences for different types of apologies. Prior work has also demonstrated that user traits impact preferences of apologies. For example, people with lower levels of extroversion, agreeableness, and conscientiousness were generally less likely to rate the agent as likeable or intelligent, regardless of the recovery strategy used \cite{cahya2021appropriate}. Similarly, individuals with lower open-mindedness tended to assign lower likeability ratings to the agent. Other individual factors that influence user preferences include social identity and technological self-efficacy. These traits influence the perception of anthropomorphic systems in turn influencing the perception of the use of apology \cite{fan2020does}.  Context is very important and can influence how people perceive apologies. For example, previous work found differences in how people react to cleaning robots in public versus private contexts \cite{babel2021development}. While there is much prior work supporting the claim that individual factors impact how users view apologies, much of this work was done in the context of chatbots that predate LLMs.

The design considerations here are different depending on whether the interaction with the chatbot is one-shot or ongoing. A customer service bot might engage with a user from a cold start, resetting to a standard configuration each time. The most that can be done with such a system is to design for either a typical or worst-case user.

In order to match user preferences and expectations, the system would need to have some pretty sophisticated information about the user. This could be collected upfront, possibly using the constructs we used to assess individual factors. Alternately, it might be extrapolated from the interaction itself.

An LLM personal assistant can support long-term and evolving interaction. This may make it possible to detect individual preferences over time based on a user's interaction history and adapt apologies accordingly. LLMs have reshaped our interactions with artificial intelligence AI systems, necessitating a need for long-term memory in some cases such as companion systems, psychological counseling \cite{zhong2024memorybank}. Recent work has proposed a Generalized User Model (GUM) that learns from any user-computer interaction, to generate confidence-weighted statements about a user’s behavior, knowledge, and preferences. GUMs can infer high-level context, like preparing for a wedding or struggling with feedback, by observing patterns across tasks. They continuously update and retrieve relevant insights to support applications like context-aware chat assistants and adaptive agents across apps \cite{shaikh2025creating}. As such models improve, they can be leveraged to offer more contextually relevant and personalized apologies.

Designs of this kind produce a new set of challenges. For instance, the `sycophantic' tendencies displayed by many current chatbots \cite{sharma2023towards,wei2023simple} might hinder their ability to accurately track user attitudes and preferences during an interaction and enact the flexible calibration required to adapt its responses to better match user attitudes and preferences.

Humans `calibrate' their responses to each other during verbal interactions when faced with an unexpected tension or mismatch in expectation. This is usually more or less explicitly signaled by one of the participants in the interaction. In the case of apologies, one party might signal that the type of apology received by the other party is not satisfactory --- perhaps because it is insufficiently remorseful, does not match the perceived amount of harm done, or misrepresents the moral record. A human making an apology is unlikely to capitulate completely. Instead, they may try to `negotiate' or `push back', asking for more information or context in order to understand why their apology was not satisfactory and assessing for themselves whether it is indeed legitimate for the other party to reject their first apology given the context. In short, ``the crucial part of an apology is the \textit{interaction}'' \cite{helmreich}. The precondition of a successful apology is shared understanding, which is often arrived at dialectically. 

Current chatbots, however, tend to assume a subordinate and servile posture towards the user almost by default, regardless of whether the nature of the breakdown or the user reaction actually warrants it \cite{magnus2025}. Chatbots are willing to `take responsibility' (and immediately apologize) for anything the user is not satisfied with, without even attempting any kind of pushback. As a consequence, the interaction may never reach the level of tension required for calibration. And active extrapolation of user background assumptions, beliefs, and preferences can hardly happen outside of this process of calibration.

\subsubsection{Attribution of Responsibility: Code, Programmers, Stakeholders}

\label{sec:attribution}

Participants in our study made meaningful distinctions between different entities potentially responsible for the error: ``the code,'' ``the programmer,'' and ``the company stakeholder.''  This was reflected in the codes generated for participant rationales regarding chatbot control and responsibility. Prior work on robot apologies has shown that when agents accept responsibility for errors (``There was a failure in my speech module''), they can better mitigate negative impacts and maintain user trust compared to those that shift blame externally \cite{correia2018exploring}. 

While prior work emphasizes the benefits of agents accepting fault, our findings suggest that users perceive responsibility within AI systems as more granular and actor-specific. Many participants did not treat ``the system'' as a monolith. Many expressed a nuanced understanding of where accountability might lie, suggesting that responsibility could be distributed across the codebase, the human developers, or the company behind the system. This delineation influenced how participants interpreted apologies: a key reason participants preferred explanatory or empathic apologies was their alignment with perceived accountability.

This raises the question: what should internal blame look like in AI-generated apologies? In our bias scenario, for example, the apology attributed the issue to biased training data. But what if the blame had instead been placed on the code itself? Or the programmer who wrote it? Or the company that deployed it? Each of these attributions could evoke different reactions. This suggests value in further inquiry into how users attribute responsibility and how those attributions interact with apology preferences, particularly in identity-relevant or ethically charged contexts.

\subsubsection{Apology Preferences for Hallucinations}
\label{sec:hallucination}
Although participants perceived unfounded fabrications in the hallucination scenario to be a serious issue, we did not find a statistically significant preference for any single apology style in this condition. While explanatory apologies were numerically preferred, followed by empathic and then rote, these differences did not reach statistical significance. This contrasts with the bias and factual error conditions, where clear preferences emerged.

One possible reason lies in the novelty and complexity of hallucinations as a failure mode. Unlike factual errors, which can often be attributed to specific gaps in knowledge, or bias-related errors, which are increasingly familiar in public discourse, hallucinations are a new and uniquely unsettling phenomenon introduced by large language models. Schneier and Sanders contrast these AI hallucinations with human tendencies \cite{schneier2025ai}: ``Humans may occasionally make seemingly random, incomprehensible, and inconsistent mistakes, but such occurrences are rare and often indicative of more serious problems.'' Beyond the field of AI, the term hallucination comes from psychology, where it denotes a distinctive kind of perception \cite{fish2009perception, macpherson2013hallucination}. Blom defines it as ``a percept, experienced by a waking individual, in the absence of an appropriate stimulus from the extracorporeal world'' \cite{blom2010dictionary}. In other words, it is an illusory experience that nevertheless feels authentic.
Researchers borrowed this label for an analogous failure mode in natural language generation: models sometimes produce text that reads fluently but is either factually ungrounded or logically incoherent \cite{ji2023survey}. Such output appears to be anchored in the given context even though that anchor is missing or unverifiable, mirroring how a clinical hallucination is hard to distinguish from genuine sensory input on first inspection. Within NLP the most widely used definition of hallucination is any generated content that is nonsensical or unfaithful to the source material \cite{filippova2020controlled, maynez2020faithfulness, parikh2020totto}. 



Participants seemed to recognize this distinctive failure mode, citing it as the one most likely to make them stop using the chatbot as reflected in their responses to the future use intention question posed across all error types. And yet, paradoxically, this did not translate into a shared sense of what kind of apology was appropriate. This may reflect a deeper uncertainty or discomfort about how to hold AI accountable for a type of mistake that feels both sophisticated and inexplicable.

Another possible explanation is that hallucinations disrupt trust on a different axis. In the factual error or bias scenarios, participants may be judging the system's moral alignment or knowledge boundaries. In hallucinations, however, they are faced with a system confidently asserting falsehoods without any obvious cue that it is doing so. The apology becomes less about emotional or explanatory repair, and more about confronting the credibility of the system's knowledge, whether users can trust that what the system says is based on anything real or verifiable.  This distinct failure mode prompted reactions from some participants, who highlighted the severity and strangeness of being presented with fabricated content:

\begin{quote}
\textit{``If it is making up an article to cite, that is a major problem. There's a difference in mixing up a time zone like the other scenario versus making an entirely fictional article up.''} ---P039
\end{quote}

\begin{quote}
\textit{``An incorrect citation is a big deal. I would probably not trust this same technology again.''} ---P071
\end{quote}

\begin{quote}
\textit{``It’s a bizarre and difficult mistake to make.''} ---P167
\end{quote}

What makes hallucinatory errors and the apologies that follow particularly compelling to study is that this kind of interaction is uncommon in human-to-human exchanges. When a person fabricates information that is easily disprovable, we might label them as a pathological or compulsive liar. As Dike et al. describe, pathological lying as a psychological diagnosis is characterized by elaborate, often impulsive falsifications that are disproportionate to any discernible gain and may even be believed by the liar themselves \cite{dike2005pathological}.  There is no true parallel between the reaction of a chatbot to a hallucination and the reaction of a pathological liar to being caught in a lie. The latter typically does not apologize or change course. Instead they deny, escalate, and gaslight or perpetuate the lie to protect their self-image \cite{danesi2024pseudology,  healy1915pathological, curtis2020pathological}. In contrast chatbots apologize freely, highlighting a reversal in repair strategy. This asymmetry may help explain the absence of a clear user preference for any specific apology type when the chatbot is confronted for hallucinating, an exchange that lacks a meaningful parallel in human interaction. 






\section{Future Work and Conclusion}
Our findings demonstrate that while explanatory apologies are broadly preferred, effective trust repair in AI-human interactions requires more than a one-size-fits-all solution. User preferences for apology style are shaped by the nature of the mistake and individual user traits. In morally charged contexts like bias, explanatory apologies risk sounding like excuses if they fail to clearly acknowledge harm. In novel failure modes like hallucinations, users struggle to agree on what kind of apology is appropriate, suggesting a mismatch between familiar human expectations and new AI behaviors. These complexities point to a broader design challenge: building systems that can not only detect the nature of an error \cite{achintalwar2024detectors,padhi2025granite}, but also respond in a manner that aligns with user expectations and contextual nuance. Achieving this level of responsiveness may require richer models of user behavior and beliefs, better calibration mechanisms, and a move away from overly sycophantic design defaults. Our study examined three apology types: rote, explanatory, and empathic guided by Smith’s theory of categorical apology \cite{smith2008}. However, other formulations, such as apologies with layered explanations or varied emotional tone, could elicit different reactions and warrant future exploration.

While our study focused on a one-shot apology, where the AI made a mistake, the user flagged it, and the system responded with a single scripted apology, real-world interactions often unfold as multi-turn, negotiated exchanges. Even in our static vignette, the character Alice’s satisfaction did not rest on explicit components of the apology, but on a general sense of resolution. This suggests that user satisfaction may emerge from a dynamic process of alignment and recalibration. Future research should examine how apology interactions evolve over time, especially in contexts where users offer corrective feedback or signal dissatisfaction, prompting the AI to adjust its tone or content accordingly.

As LLMs increasingly mediate sensitive or high-stakes interactions, understanding how apology functions as a social repair strategy becomes essential for developing trustworthy AI systems. Yet trust repair may not always hinge on delivering an apology at all. Some participants in our study found the chatbot responses overly long or filled with unnecessary detail, hinting that, in certain situations, no apology, or a different form of acknowledgment, might have been preferred. Future research should explore these boundaries by incorporating ``no-apology'' conditions and examining when users prefer alternative forms of repair. As we move toward AI systems that not only act but also account for their actions, designing repair strategies that are contextually appropriate and user-sensitive will be central to fostering meaningful and responsible human-AI interaction. 


\bibliographystyle{ACM-Reference-Format}
\bibliography{main}
\end{document}